\documentclass[%
reprint,
 superscriptaddress,
 preprintnumbers,
 nofootinbib,
 amsmath,amssymb,
 pra,final,
 showpacs,twocolumn,
 aps]{revtex4-2}
\usepackage[hyperindex,breaklinks]{hyperref}
\hypersetup{
        colorlinks=true,
        linkcolor=blue,
        filecolor=magenta,
        urlcolor=blue,
        citecolor=magenta
}

\usepackage[utf8]{inputenc}
\usepackage[english]{babel}

\usepackage{epsfig}
\usepackage{graphicx}
\usepackage{amsfonts}
\usepackage{enumerate}
\usepackage{mathrsfs}

\usepackage{amsmath,amssymb}
\usepackage{latexsym}
\usepackage{slashed}
\usepackage{soul}

\usepackage{bbm}
\usepackage{dcolumn}
\newcolumntype{+}{D{+}{\,\pm\,}{-3,3}}

\newcommand{\be}{\begin{equation}}
\newcommand{\ee}{\end{equation}}
\newcommand{\ba}{\begin{eqnarray}}
\newcommand{\ea}{\end{eqnarray}}
\newcommand{\bs}{\begin{subequations}}
\newcommand{\es}{\end{subequations}}
\newcommand{\no}{\nonumber\\}

\graphicspath{ {figures/} }

\begin{document}

\title{\LARGE On the extension of the SM through a scalar quadruplet}
\author{Darius~Jur\v{c}iukonis}
\email{darius.jurciukonis@tfai.vu.lt}
\affiliation{Vilnius University, Institute of Theoretical Physics and Astronomy, \\ Saul\.etekio~av.~3, Vilnius 10257, Lithuania}
\author{Lu\'\i s~Lavoura}
\email{balio@cftp.tecnico.ulisboa.pt}
\affiliation{Universidade de Lisboa, Instituto Superior T\'ecnico, CFTP, \\
 Av.~Rovisco~Pais~1, 1049-001~Lisboa, Portugal}

\begin{abstract}
  We consider the extension of the Standard Model (SM)
  through a scalar quadruplet with hypercharge either $1/2$ or $3/2$;
  in the first case,
  we assume $CP$ invariance of the scalar potential (SP).
  We write down the unitarity conditions on the SP.
  We use a partly numerical method
  to find the bounded-from-below conditions on the SP.
  We determine the masses of the new scalars of the model.
  We compute the three- and four-Higgs couplings
  and compare them to the ones of the SM.
\end{abstract}

\maketitle

\section{Introduction}
\label{sec:intro}

In this paper we consider
the Standard Model (SM) extended by an $SU(2)$ quadruplet of scalars.
The scalar sector of this extension of the SM consists of
\be
\label{multiplets1}
\Phi = \left( \begin{array}{c} a \\ b \end{array} \right)
\quad \mbox{and} \quad
\Psi = \left( \begin{array}{c} c \\ d \\ e \\ f \end{array} \right),
\ee
where $\Phi$ is the Higgs doublet of the SM,
$\Psi$ is the additional quadruplet,
and $a, \ldots, f$ are complex scalar fields.
Let $I$ be the third component of weak isospin;
then,
$c$ has $I=3/2$,
$a$ and $d$ have $I=1/2$,
$b$ and $e$ have $I=-1/2$,
and $f$ has $I=-3/2$.
Let $Y$ be the weak hypercharge\footnote{In our normalization
the electric charge is $I + Y$.};
both $\Phi$ and $\Psi$ have $Y = 1/2$.
Therefore,
the fields $b$ and $e$ are electrically neutral.\footnote{In Appendix~\ref{Y32}
we treat the case where the hypercharge of $\Psi$ is $3/2$.
In that case,
the neutral fields are $b$ and $f$.}

We denote
\be
A \equiv \left| a \right|^2, \
\ldots, \
F \equiv \left| f \right|^2.
\ee

The following products of representations are $SU(2)$ triplets with $Y = 1$:
\bs
\label{tri}
\ba
\left( \Psi \otimes \Phi \right)_\mathbf{3} &=&
\left( \begin{array}{c}
  \sqrt{3}\, bc - ad \\ \sqrt{2} \left( bd - ae \right) \\ be - \sqrt{3}\, af 
\end{array} \right),
\label{10c}
\\
\left( \Phi \otimes \Phi \right)_\mathbf{3} &=&
\left( \begin{array}{c} a^2 \\ \sqrt{2}\, ab \\ b^2 \end{array} \right),
\label{T1} \\
\left( \Psi \otimes \Psi \right)_\mathbf{3} &=&
\left( \begin{array}{c}
  d^2 - \sqrt{3}\, ce \\ \left( d e - 3 c f \right) \left/ \sqrt{2} \right. \\
  e^2 - \sqrt{3}\, df
\end{array} \right).
\label{10b}
\ea
\es
(We have chosen the normalizations of the triplets in a convenient way.)
One generates gauge-invariant quantities
by multiplying each triplet by the Hermitian conjugate of another one:
\bs
\label{prods}
\begin{align}
& \left( \Psi \otimes \Phi \right)_\mathbf{3}^\dagger \,
\left( \Phi \otimes \Phi \right)_\mathbf{3}
\equiv \mathcal{F}_6 
\\  &\phantom{WW} =  \left( B - 2 A \right) b e^\ast
+ \left( 2 B -  A \right) a d^\ast \no 
&\phantom{WW=} + \sqrt{3} \left( a^2 b^\ast c^\ast - b^2 a^\ast f^\ast \right),
\label{F6}
\\
& \left( \Psi \otimes \Phi \right)_\mathbf{3}^\dagger \,
\left( \Psi \otimes \Psi \right)_\mathbf{3}
\equiv \mathcal{F}_7
\\ &\phantom{WW} =
\left( D + E \right) \left( b^\ast e - a^\ast d \right) \no
&\phantom{WW=} 
+ \sqrt{3} \left( a^\ast d^\ast c e + b^\ast c^\ast d^2
- b^\ast e^\ast d f - a^\ast f^\ast e^2 \right) \no
&\phantom{WW=} 
+ 3 \left( a^\ast e^\ast c f - b^\ast d^\ast c f + F a^\ast d - C b^\ast e \right),
\label{F7}
\\
& \left( \Phi \otimes \Phi \right)_\mathbf{3}^\dagger \,
\left( \Psi \otimes \Psi \right)_\mathbf{3}
\equiv \mathcal{F}_8
\\ &\phantom{WW} = {b^\ast}^2 e^2 + {a^\ast}^2 d^2 +  a^\ast b^\ast d e \no
&\phantom{WW=} - \sqrt{3} \left( {b^\ast}^2 d f + {a^\ast}^2 c e \right)
- 3 a^\ast b^\ast c f.
\label{F8}
\end{align}
\es

The scalar potential (SP)
has a quadratic part $V_2$ and a quartic part $V_4$:
\be
V = V_2 + V_4,
\label{potV}
\vspace*{-1mm}
\ee
where
\bs
\ba
V_2 &=& \mu_1^2 F_1 + \mu_2^2 F_2,
\\
F_1 &\equiv& A + B, \label{F1}
\\
F_2 &\equiv& C + D + E + F, \label{F2}
\ea
\es
and
\ba
\label{V4}
V_4 &=& \sum_{p=1}^2\, \frac{\lambda_p}{2}\, F_p^2
+ \lambda_3 F_1 F_2
+ \lambda_4 F_4
+ \lambda_5 F_5 \no
&&+ \left( \sum_{p=6}^8\, \frac{\lambda_p}{2}\, \mathcal{F}_p
+ \mathrm{H.c.} \right).
\ea
In Eq.~\eqref{V4},
$\lambda_1$,
$\lambda_2$,
$\lambda_3$,
$\lambda_4$,
and $\lambda_5$ are real coefficients,
while $\lambda_6$,
$\lambda_7$,
and $\lambda_8$ are (in general) complex coefficients.
Furthermore,\footnote{$F_4$ and $F_5$ have been defined
in Ref.~\cite{ourrecent} with extra factors $1/4$ and $1/5$,
respectively.}
\bs
\begin{align}
F_4 \equiv& \,\left( A - B \right) \left( 3 C + D - E - 3 F \right) \no
&+ \left[ 2 \sqrt{3}\left( a f b^\ast e^\ast + a d b^\ast c^\ast \right)
+ 4 a e b^\ast d^\ast + \mathrm{H.c.} \right],\hspace*{-1mm}
\label{F4}
\end{align}
\begin{align}
F_5 \equiv& \, D E + 2 \left( D^2 + E^2 \right) + 6 \left( C E + D F \right)
+ 9 C F \no 
&- \left[ 2 \sqrt{3}\left( d^2 c^\ast e^\ast + e^2 d^\ast f^\ast \right)
+ 3 c f d^\ast e^\ast + \mathrm{H.c.} \right]\hspace*{-1mm}
\label{F5}
\end{align}
\es
are $SU(2) \times U(1)$-invariant quantities.

In this paper we \emph{assume} $\lambda_6$,
$\lambda_7$,
and $\lambda_8$ to be \emph{real},
which is the assumption of $CP$ invariance of the SP.
If those three coefficients are not real,
then the minimization of $V_4$ that we
perform in subsection~\ref{mini} of Appendix~\ref{BFBapp} is not possible.

The electrically neutral fields $b$ and $e$
develop vacuum expectation values (VEVs) $u$ and $t$,
respectively.
Note that,
because there are in $V_4$ terms $B b e^*$ and $E b e^*$
(and their Hermitian conjugates),
$u \neq 0$ implies $t \neq 0$ and vice-versa.
We denote
\be
U = \left| u \right|^2, \quad T = \left| t \right|^2, \quad
\theta = \arg{\left( u t^\ast \right)}.
\label{UTtheta}
\ee
The masses $m_W$ and $m_Z$ of the $W^\pm$ and $Z^0$ gauge bosons,
respectively,
are given by~\cite{albergaria}
\bs
\ba
m_W^2 &=& g^2\, \frac{U + 7 T}{2},
\\
m_Z^2 &=& \frac{g^2}{c_W^2}\, \frac{U + T}{2}.
\ea
\es
Since
\be
G_F = \frac{g^2}{4 \sqrt{2} m_W^2}
\ee
and
\be
c_W^2 = 1 - \frac{\pi \alpha_\mathrm{em}}{\sqrt{2} G_F m_W^2},
\ee
one has
\bs
\label{3}
\ba
U + 7 T &=& \frac{1}{2 \sqrt{2} G_F}, \label{6a}
\\
U + T &=& \frac{m_Z^2 \left( \sqrt{2} G_F m_W^2
  - \pi \alpha_\mathrm{em} \right)}{\left( 2 G_F m_W^2 \right)^2}.
\ea
\es
Equations~\eqref{3} allow one to determine $U$ and $T$
as functions of the \emph{input observables}
$G_F$ (the Fermi coupling constant),
$\alpha_\mathrm{em}$ (the fine-structure constant),
$m_W^2$,
and $m_Z^2$.\footnote{In this extension of the SM,
$m_W^2 \left/ \left( c_W^2 m_Z^2 \right) \right. \neq 1$ at the tree level.
This implies that the renormalization of the gauge sector of this model
requires four input observables instead of just three.
For the same reason,
it is \emph{not} possible to define oblique parameters for this extension
of the SM~\cite{drauksas}.}
Since $U + 7 T \ge U + T$,
those input observables must be chosen so that
\ba
\label{condinput}
\sqrt{2} G_F m_W^4 &\ge& m_Z^2 \left( \sqrt{2} G_F m_W^2 - \pi \alpha_\mathrm{em}
\right).
\ea
%
This agrees with a recent measurement by the CDF Collaboration~\cite{CDF}
that produced $m_W$ larger than the one expected in the context of the SM.

Note that
\be
v_\mathrm{SM} \equiv \sqrt{U + 7 T} = \sqrt{\frac{1}{2 \sqrt{2}\, G_F}}
= 174.104\,\mathrm{GeV}
\label{vSM}
\ee
is the value of the VEV of the Higgs doublet in the SM.

\begin{widetext}

\begin{figure}[t]
\begin{center}
\includegraphics[width=0.9\textwidth]{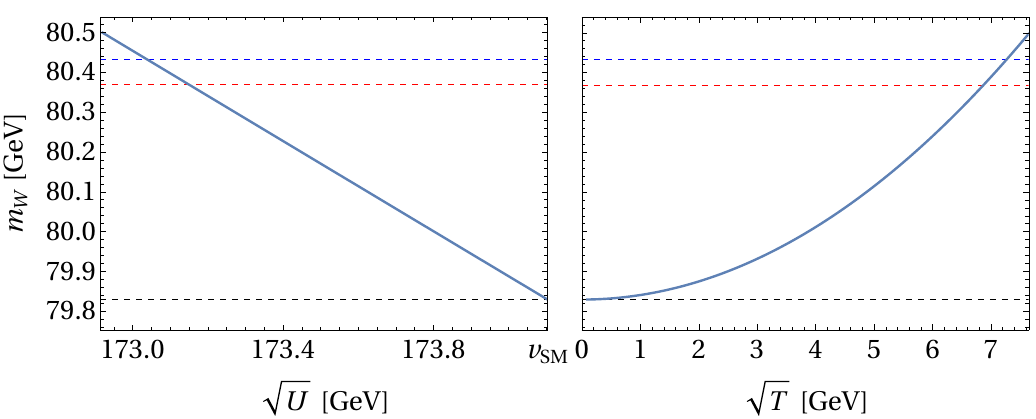}
\end{center}
\caption{The dependence of the VEVs $\sqrt{U}$ (left panel)
  and $\sqrt{T}$ (right panel)
  from the mass $m_W$ of the gauge boson $W^\pm$,
  according to Eqs.~\eqref{3}.
  The inputted values of $m_Z$,
  $\alpha_\mathrm{em}$,
  and $G_F$ are given in Eqs.~\eqref{theinputs}.
  The black dashed line indicates the
  (tree-level)
  value $m_W^\mathrm{SM} = 79.829$\,GeV of $m_W$ in the SM.
  The red dashed line indicates the most recent experimental world average
  value of $m_W$~\cite{pdg};
  the blue dashed line indicates the value of $m_W$
  recently divulged by the CDF experiment~\cite{CDF}.
  \label{fig:VEVs_Y12}}
\end{figure}

\end{widetext}

In this paper we always use
the input values~\cite{pdg}
\bs
\label{theinputs}
\ba
m_Z &=& 91.188,0\,\mathrm{GeV},
\label{inputmZ} \\
\alpha_\mathrm{em} &=& \frac{1}{127.930},
\label{inputalpha} \\
G_F &=& 1.166,378,8 \times 10^{-5}\,\mathrm{GeV}^{-2}.
\label{inputGF}
\ea
\es
Then,
the relation between the input $m_W$ and the VEVs $\sqrt{U}$ and $\sqrt{T}$
is depicted in Fig.~\ref{fig:VEVs_Y12}.
One sees that any $m_W$ slightly larger than the one predicted by the SM
may be explained by a value of $T$ much smaller than the one of $U$.

In this model there is one double-charged scalar,
two single-charged scalars,
and three neutral scalars.
One of the latter is supposed to be the particle $h$
with squared mass
\be
\label{125}
M_h = \left( 125\,\mathrm{GeV} \right)^2
\ee
that has been observed at the LHC.
That particle will in general enjoy both cubic and quartic self-interactions,
given by the Lagrangian terms
\be
\mathcal{L} = \cdots - g_3 h^3 - g_4 h^4.
\ee
In the SM,
$g_3$ and $g_4$ may be predicted to be
\bs
\label{g34sm}
\ba
g_3^\mathrm{SM} = \frac{M_h}{2 \sqrt{2}\, v_\mathrm{SM}}
&=& 31.729,8\,\mathrm{GeV},
\\
g_4^\mathrm{SM} = \frac{M_h}{16\, v_\mathrm{SM}^2} &=& 0.032,217,
\ea
\es
respectively.
In extensions of the SM,
$g_3$ and $g_4$ cannot be exactly predicted;
new heavy scalars may indirectly influence them in diverse ways.
Thus,
studying the Higgs self-couplings
could provide valuable insights into New Physics.
In this paper we investigate,
among other things,
the range of possible values of $g_3$ and $g_4$
in the quadruplet extension of the SM.
Notice however that \textit{a priori}\/
we do not expect $g_3$ and $g_4$ to differ much from $g_3^\mathrm{SM}$
and $g_4^\mathrm{SM}$,
respectively,
because $T \ll U$ implies that the quadruplet model
remains very close to the SM.

Experimentally,
$g_3$ is largely unconstrained;
for example,
ATLAS reports $-0.4 \le g_3 \left/ g_3^\mathrm{SM} \right. \le 6.3$
at the 95\% confidence level (C.L.)~\cite{ATLAS:2022jtk}. 
At the high-luminosity LHC,
the projected sensitivity is expected to be
$0.1 \le g_3 \left/ g_3^\mathrm{SM} \right. \le 2.3$
at 95\% C.L.~\cite{Cepeda:2019klc} or,
more recently,
$0.5 \le g_3 \left/ g_3^\mathrm{SM} \right. \le 1.6$
at 68\% C.L.~\cite{ATLAS:2022faz}.
Future colliders may possibly constrain $g_3$
to within 10\% of $g_3^\mathrm{SM}$~\cite{Cepeda:2019klc,Goncalves:2018qas}. 

The cubic and quartic Higgs self-couplings
in the two-Higgs-doublet model have been studied quite extensively
both at the tree level~\cite{Haber:2006ue,
  Osland:2008aw, Celis:2013rcs, Ginzburg:2015yva, we}
and accounting for loop corrections~\cite{Kanemura:2004mg,
  Braathen:2019pxr, Braathen:2019zoh, Bahl:2022jnx, Heinemeyer:2024hxa}.
There is also much research done on the Inert Doublet Model,
with Refs.~\cite{Arhrib:2015hoa,
  Braathen:2019pxr, Braathen:2019zoh, Falaki:2023tyd}
claiming that the deviations of the cubic Higgs self-coupling
from the SM value can exceed 100\%.
The cubic Higgs self-couplings in the extension of the SM
through a scalar triplet has been studied
in Refs.~\cite{Chen:2008jg, Aoki:2012jj, Dawson:2017vgm}, 
while the phenomenology of the SM extended
through a scalar quadruplet has been studied
in Refs.~\cite{AbdusSalam:2013eya, Dawson:2017vgm};
it remains relevant today,
with recent studies on scalar quadruplets
in Refs.~\cite{roma1, roma2, kannikerecent}.

In this paper we impose various constraints on the SP of the model,
in particular the unitarity (UNI)
constraints described in section~\ref{sec:UNI}
and the bounded-from-below (BFB)
constraints described in section~\ref{sec:BFB};
both constrain only the parameters of $V_4$.
In section~\ref{sec:unique} we present the three possible vacua of the model,
namely the two $CP$-conserving vacua and one spontaneously $CP$-breaking vacuum.
In section~\ref{sec:scalarmasses} we compute the masses of all the scalars.
In section~\ref{sec:g3g4} we compute $g_3$ and $g_4$.
In section~\ref{sec:procedure} we outline the fitting procedure
that we have utilized and in section~\ref{sec:results} we give our
numerical findings.
A short discussion of our conclusions is made in section~\ref{sec:conclusions}.

Appendix~\ref{BFBapp} consists of a detailed derivation of the BFB conditions.
In Appendix~\ref{uniq} we discuss constraints on the potential
derived from the requirement that our preferred vacuum state
is indeed the global minimum of $V$.

The $SU(2)$ quadruplet $\Psi$ can interact with the Higgs doublet $\Phi$
in different ways,
depending on its hypercharge $Y$.
The phenomenologically interesting possibilities are $Y = 1/2$,
discussed in the core of this paper,
and $Y = 3/2$,
discussed only in Appendix~\ref{Y32}.
However,
in the case $Y = 3/2$ the $W^\pm$ mass must obey condition~\eqref{cond_Y32},
which is the opposite of condition~\eqref{condinput}
of the $Y = 1/2$ case---and,
moreover,
tends to contradict the experimental measurements of $m_W$.

\section{Unitarity (UNI) conditions}
\label{sec:UNI}

We have derived the UNI conditions on $V_4$
by running an appropriate {\tt Mathematica} code
that we have written ourselves.
There are only two scattering channels
to be considered\footnote{All the other channels
yield UNI conditions that reproduce
some of those that we have derived through these two channels.}:
\begin{enumerate}
\item The channel of the two-field states with $I = Y = 1$,
  \textit{viz.}\ the five states $a a$,
  $d d$,
  $a d$,
  $b c$,
  and $c e$.
\item The channel of the two-field states with $I = Y = 0$,
  \textit{viz.}\ the ten states $a a^\ast$,
  $b b^\ast$,
  $c c^\ast$,
  $d d^\ast$,
  $e e^\ast$,
  $f f^\ast$,
  $a e^\ast$,
  $a^\ast e$,
  $b d^\ast$,
  and $b^\ast d$.
\end{enumerate}
The UNI conditions that we have obtained from these two channels are:
\bs
\ba
\left| \lambda_2 \right| &<& M, \\
\left| \lambda_2 + 9 \lambda_5 \right| &<& M, \\
\left| \lambda_3 + 3 \lambda_4 \right| &<& M, \\
\left| 2 \lambda_3 - 6 \lambda_4 - 3 \lambda_8
\right| &<& 2 M, \\
\left| 2 \lambda_3 + 10 \lambda_4 - 5 \lambda_8
\right| &<& 2 M, \\
\left| 3 \lambda_1 + Q_1 \right| 
+ \sqrt{\left( 3 \lambda_1 - Q_1 \right)^2
  + 32 \lambda_3^2} &<& 2 M,
\\
\left| 2 \lambda_2 + 6 \lambda_5 + Q_2 \right|
& & \no
+ \sqrt{\left( 2 \lambda_2 + 6 \lambda_5 - Q_2 \right)^2
  + 144 \lambda_7^2} &<& 4 M,
\ea
\es
where $Q_1 = 5 \lambda_2 + 15 \lambda_5$, $Q_2 = 2 \lambda_3 - 6 \lambda_4 + 3 \lambda_8$, and we take $M = 8 \pi$.
Besides,
the moduli of the eigenvalues of the matrices
\bs
\label{mat}
\ba
&
\left( \begin{array}{ccc}
  \lambda_1 & - \sqrt{5/2}\, \lambda_8 & \sqrt{2}\, \lambda_6 \\
  - \sqrt{5/2}\, \lambda_8 & \lambda_2 + 10 \lambda_5 &
  - \sqrt{5}\, \lambda_7 \\
  \sqrt{2}\, \lambda_6 & - \sqrt{5}\, \lambda_7 & \lambda_3 - 5 \lambda_4
\end{array} \right), &
\\
&
\left( \begin{array}{ccc}
  \lambda_1 & 2 \sqrt{10}\, \lambda_4 & - 2 \sqrt{2}\, \lambda_6 \\
  2 \sqrt{10}\, \lambda_4 & \lambda_2 - 11 \lambda_5 &
  - \sqrt{5}\, \lambda_7 \\
  - 2 \sqrt{2}\, \lambda_6 & - \sqrt{5}\, \lambda_7 &
  \lambda_3 + 5 \lambda_4 + 5 \lambda_8 / 2
\end{array} \right)
&
\ea
\es
must be smaller than $M$.

When $\lambda_6 = \lambda_7 = \lambda_8 = 0$,
these conditions reduce to conditions~(25),
(83),
(87),
(88),
(89),
and~(A2) of Ref.~\cite{ourrecent}.\footnote{Note that
the $\lambda_4$ of this paper is $\lambda_4 / 4$ in Ref.~\cite{ourrecent}
and the $\lambda_5$ of this paper is $\lambda_5 / 5$ in Ref.~\cite{ourrecent}.}

\section{Bounded-from-below (BFB) conditions}
\label{sec:BFB}

The derivation of the necessary and sufficient BFB conditions
is done in Appendix~\ref{BFBapp};
we make here just a practical summary of them.
We firstly check the \emph{necessary} conditions
\bs
\label{iop}
\ba
\lambda_1 &>& 0, \\
\lambda_2 &>& 0, \\
2 \lambda_2 + 9 \lambda_5 &>& 0.
\ea
\es
We then make a scan over $x$ from $0$ to $1$,
with another parameter $y$ given by either $y=0$
or $y = \left( 9/5 \right) x \left( 1 - x \right)$.
\begin{itemize}
\item If $\lambda_6 \lambda_7 \lambda_8 < 0$,
  then for every $x$ and $y$ we define
  \bs
  \label{c40}
  \ba
  c_4 &=& \frac{\lambda_1}{2},
  \\
  c_0 &=& \frac{\lambda_2}{2} + 5 \lambda_5 y,
  \ea
  \es
  and
  \bs
  \ba
  c_3 &=& - \sqrt{3 x} \left| \lambda_6 \right|, \\
  c_2 &=& \lambda_3 + 3 \lambda_4 \left( 1 - 2 x \right)
  - \sqrt{\frac{5 y}{2}} \left| \lambda_8 \right|, \\
  c_1 &=& - \sqrt{\frac{15 x y}{2}} \left| \lambda_7 \right|,
  \ea
  \es
  \bs
  \label{alfabetagama}
  \ba
  \psi_3 &\equiv& \frac{c_3}{\sqrt[4]{c_4^3\, c_0}},
  \\
  \psi_2 &\equiv& \frac{c_2}{\sqrt{c_4\, c_0}},
  \\
  \psi_1 &\equiv& \frac{c_1}{\sqrt[4]{c_4\, c_0^3}},
  \ea
  \es
  \bs
  \label{DeltaLambdas}
  \ba
  \Delta &\equiv&
  4 \left( \psi_2^2 - 3 \psi_1 \psi_3 + 12 \right)^3 
  \left( 72 \psi_2 + 9 \psi_1 \psi_2 \psi_3 \right. \no  
  && \left. - 2 \psi_2^3 - 27 \psi_1^2
  - 27 \psi_3^2 \right)^2, \\
  \Lambda_1 &\equiv& \left( \psi_1 - \psi_3 \right)^2
  - 16 \left( \psi_1 + \psi_2 + \psi_3 + 2 \right),
  \\
  \Lambda_2 &\equiv& \left( \psi_1 - \psi_3 \right)^2
  - 4 \left( \psi_2 + 2 \right) \left( 4
  + \frac{\psi_1 + \psi_3}{\sqrt{\psi_2 - 2}} \right). \hspace*{3mm}
  \ea
  \es
  For BFB we require that
  \bs
  \label{poss}
  \ba
  \label{poss1}
  \mbox{either}\;\;  & &
  \Delta \le 0 \hspace*{2mm} \mbox{and} \hspace*{2mm} \psi_1 + \psi_3 > 0,
  \\
  \label{poss2}
  \mbox{or} \;\; & &
  -2 \le \psi_2 \le 6, \hspace*{2mm} \Delta \ge 0, \hspace*{2mm} \mbox{and} \hspace*{2mm}
  \Lambda_1 \le 0, \hspace*{2mm}
  \\
  \label{poss3}
  \mbox{or} \;\; & &
  \psi_2 > 6, \hspace*{2mm} \psi_3 > 0, \hspace*{2mm} \mbox{and} \quad \psi_1 > 0,
  \\
  \label{poss4}
  \mbox{or} \;\; & & \psi_2 > 6, \hspace*{2mm} \Delta \ge 0, \hspace*{2mm} \mbox{and}
  \quad \Lambda_2 \le 0.
  \ea
  \es
  If condition~\eqref{poss} holds for every $x$,
  and for both $y=0$ and $y = \left( 9/5 \right) x \left( 1 - x \right)$,
  then $V_4$ is BFB.
\item If $\lambda_6 \lambda_7 \lambda_8 > 0$,
  then for every $x$ and $y$ we use Eqs.~\eqref{c40} together with
  \bs
  \ba
  c_3 &=& \left( -1 \right)^p \sqrt{3 x} \left| \lambda_6 \right|,
  \\
  c_2 &=& \lambda_3 + 3 \lambda_4 \left( 1 - 2 x \right)
  + \left( -1 \right)^{p+q} \sqrt{\frac{5 y}{2}} \left| \lambda_8 \right|,
  \\
  c_1 &=& \left( -1 \right)^q \sqrt{\frac{15 x y}{2}} \left| \lambda_7 \right|,
  \ea
  \es
  for three cases:
  (1) $p=q=1$,
  (2) $p=1$ and $q=0$,
  and (3) $p=0$ and $q=1$.
  We define the quantities in Eqs.~\eqref{alfabetagama} and~\eqref{DeltaLambdas}
  and we require that condition~\eqref{poss} holds---it must hold
  for all three cases and for every $x$ and $y$ in order that $V_4$ is BFB.
  We then define
  \bs
  \label{theLs}
  \ba
  l_4 &=& - \left| \lambda_6 \lambda_8 \right|^4,
  \label{L4} \\
  l_3 &=& 6 \left| \lambda_6^2 \lambda_7 \lambda_8 \right|^2 x,
  \\
  l_2 &=& 5 \left| \lambda_6 \lambda_7 \lambda_8^2 \right|^2 y
  - 9 \left| \lambda_6 \lambda_7 \right|^4 x^2,
  \\
  l_1 &=& 15 \left| \lambda_6 \lambda_7^2 \lambda_8 \right|^2 x y,
  \\
  l_0 &=& - \frac{25}{4} \left| \lambda_7 \lambda_8 \right|^4 y^2,
  \label{L0}
  \ea
  \es
  and we look for real and positive roots of equation
  \be
  \label{lll}
  l_4 r^4 + l_3 r^3 + l_2 r^2 + l_1 r + l_0 = 0
  \ee
  for $r$.
  If that equation has no real and positive roots,
  then we do not need to proceed any further;
  otherwise,
  we need to perform extra tests:
  \begin{enumerate}
  \item If Eq.~\eqref{lll} has only two real and positive roots,
    say $r_1$ and $r_2$,
    then we define
    \bs
    \label{kfjodp1}
    \ba
    p_2 &=& \frac{1}{2} \left( \lambda_1
    - \left| \frac{\lambda_6 \lambda_8}{\lambda_7} \right| \right),
    \\
    p_1 &=& \lambda_3 + 3 \lambda_4 - \left( 6 \lambda_4 +
    \left| \frac{3 \lambda_6 \lambda_7}{2 \lambda_8} \right| \right) x,
    \\
    p_0 &=& \frac{\lambda_2}{2} + 5 \left( \lambda_5
    - \left| \frac{\lambda_7 \lambda_8}{4 \lambda_6} \right| \right) y,
    \ea
    \es
    \bs
    \label{kfjodp2}
    \ba
    R_1 &\equiv& p_2 r_1^2 + p_1 r_1 + p_0,
    \\
    R_2 &\equiv& p_2 r_2^2 + p_1 r_2 + p_0,
    \\
    R_3 &\equiv& p_2 r_1 r_2 + p_1\, \frac{r_1 + r_2}{2} + p_0.
    \ea
    \es
    For $V_4$ to be BFB we require that
    \bs
    \label{kfjodp3}
    \ba
    R_1 &\ge& 0,
    \\
    R_2 &\ge& 0,
    \\
    R_3 + \sqrt{R_1 R_2} &\ge& 0.
    \ea
    \es
  \item If Eq.~\eqref{lll} has four real and positive roots $r_{1,2,3,4}$,
    then we firstly order them as $r_1 < r_2 < r_3 < r_4$.
    Secondly,
    we use the definitions~\eqref{kfjodp1} and~\eqref{kfjodp2}
    and we require that conditions~\eqref{kfjodp3} hold.
    Thirdly,
    we use again the definitions~\eqref{kfjodp1} and~\eqref{kfjodp2}
    but with $r_1 \to r_3$ and $r_2 \to r_4$,
    and we once again require that conditions~\eqref{kfjodp3} hold.
  \end{enumerate}
\end{itemize}
With this procedure---and provided the scan over $x$ is sufficiently
fine-grained---one obtains the \emph{necessary and sufficient}
BFB conditions on $V_4$.

\section{$CP$-conserving and -breaking vacua}
\label{sec:unique}

The terms of the potential $V$
that only depend on the electrically neutral fields $b$ and $e$
form
\ba
\label{vacpot}
V_{b,e} &=& \mu_1^2 B
+ \mu_2^2 E
+ \frac{\lambda_1}{2} B^2
+ \left( \frac{\lambda_2}{2} + 2 \lambda_5 \right) E^2
\no & &
+ \left( \lambda_3 + \lambda_4 \right) B E
+ \left( \frac{\lambda_6}{2} B + \frac{\lambda_7}{2} E \right)
\left( b^\ast e + b e^\ast \right)
\no & &
+ \frac{\lambda_8}{2}
\left[ \left( b^\ast e \right)^2 + \left( b e^\ast \right)^2 \right].
\ea
Hence,
\ba
\label{n7m8}
V_0 &\equiv& \left\langle 0 \left| V_{b,e} \right| 0 \right\rangle \no
&=&
\mu_1^2 U + \mu_2^2 T
+ \frac{\lambda_1}{2}\, U^2
+ \left( \frac{\lambda_2}{2} + 2 \lambda_5 \right) T^2
\no & &
+ \left( \lambda_3 + \lambda_4 \right) U T
+ \Lambda \sqrt{U T} \cos{\theta}
\no & &
+ \lambda_8 U T \cos{\left( 2 \theta \right)},
\ea
where
\be
\label{Lambda}
\Lambda \equiv \lambda_6 U + \lambda_7 T.
\ee
The extremum equations for $V_0$ in Eq.~\eqref{n7m8} are
\bs
\label{80}
\ba
0 = \frac{\partial V_0}{\partial U} &=& \mu_1^2 + \lambda_1 U
+ \left( \lambda_3 + \lambda_4 \right) T
+ \lambda_8 T \cos{\left( 2 \theta \right)} 
\no & &
+ \left( \frac{3 \lambda_6}{2}
+ \frac{\lambda_7 T}{2 U}\ \right) \sqrt{U T} \cos{\theta},
\\
0 = \frac{\partial V_0}{\partial T} &=& \mu_2^2
+ \left( \lambda_2 + 4 \lambda_5 \right) T
+ \left( \lambda_3 + \lambda_4 \right) U
\no & &
+ \left( \frac{\lambda_6 U}{2 T}
+ \frac{3 \lambda_7}{2} \right) \sqrt{U T} \cos{\theta}
\no & &
+ \lambda_8 U \cos{\left( 2 \theta \right)},
\\
0 = \frac{\partial V_0}{\partial \theta} &=&
- \left( \Lambda \sqrt{U T}
+ 4 \lambda_8 U T \cos{\theta} \right) \sin{\theta}.
\ea
\es
They have a spontaneously $CP$-breaking solution\footnote{The scalar potential
with real $\lambda_6$,
$\lambda_7$,
and $\lambda_8$ preserves the symmetry $CP$.
The vacuum~\eqref{stab} breaks that symmetry.
The vacua~\eqref{stab0} and~\eqref{stabpi} conserve it.}
\bs
\label{stab}
\begin{align}
\label{costeta}
\cos{\theta} = & \, \frac{- \Lambda}{4 \lambda_8 \sqrt{U T}},
\\
\mu_1^2 = &
\left( - \lambda_1 + \frac{\lambda_6^2}{4 \lambda_8} \right) U
+ \left( - \lambda_3 - \lambda_4 + \lambda_8
+ \frac{\lambda_6 \lambda_7}{4 \lambda_8} \right) T ,
\label{69b}
\\
\mu_2^2 =&
\left( - \lambda_2 - 4 \lambda_5 + \frac{\lambda_7^2}{4 \lambda_8}
\right) T
\no &
+ \left( - \lambda_3 - \lambda_4 + \lambda_8
+ \frac{\lambda_6 \lambda_7}{4 \lambda_8} \right) U,
\label{69e}
\\
V_0 =&
- \frac{\lambda_1}{2}\, U^2
+ \left( - \frac{\lambda_2}{2} - 2 \lambda_5 \right) T^2
\no &
+ \left( \lambda_8 - \lambda_3 - \lambda_4 \right) U T
+ \frac{\Lambda^2}{8 \lambda_8}.
\label{v02}
\end{align}
\es
The solution~\eqref{costeta} for $\theta$
exists and yields the minimum of $V_0$ if and only if
\be
\label{ine}
\lambda_8 > 0 \quad \mbox{and} \quad
- 4 \lambda_8 \sqrt{U T} < \Lambda < 4 \lambda_8 \sqrt{U T}.
\ee
Equations~\eqref{69b} and~\eqref{69e}
may be inverted to give $U$ and $T$
as \emph{unique} functions of the parameters of the potential.

Equations~\eqref{80} have two other solutions:
\bs
\label{stab0}
\begin{align}
\theta =& \;0,
\\
\mu_1^2 =& - \lambda_1 U - \left( \lambda_3 + \lambda_4 + \lambda_8 \right) T
\no &
- \frac{3 \lambda_6}{2}\, \sqrt{U T}
- \frac{\lambda_7 T}{2 U}\, \sqrt{U T},
\label{os1} \\
\mu_2^2 =& - \left( \lambda_2 + 4 \lambda_5 \right) T
- \left( \lambda_3 + \lambda_4 + \lambda_8 \right) U
\no &
- \frac{\lambda_6 U}{2 T}\, \sqrt{U T}
- \frac{3 \lambda_7}{2}\, \sqrt{U T},
\label{os2} \\
V_0 =&
- \frac{\lambda_1}{2}\, U^2
- \left( \frac{\lambda_2}{2} + 2 \lambda_5 \right) T^2
\no &
- \left( \lambda_3 + \lambda_4 + \lambda_8 \right) U T
- \Lambda \sqrt{U T};
\end{align}
\es
and
\bs
\label{stabpi}
\begin{align}
\theta =& \;\pi,
\\
\mu_1^2 =& - \lambda_1 U
- \left( \lambda_3 + \lambda_4 + \lambda_8 \right) T
\no &
+ \frac{3 \lambda_6}{2}\, \sqrt{U T}
+ \frac{\lambda_7 T}{2 U}\, \sqrt{U T},
\label{os3} \\
\mu_2^2 =& - \left( \lambda_2 + 4 \lambda_5 \right) T
- \left( \lambda_3 + \lambda_4 + \lambda_8 \right) U
\no &
+ \frac{\lambda_6 U}{2 T}\, \sqrt{U T}
+ \frac{3 \lambda_7}{2}\, \sqrt{U T},
\label{os4} \\
V_0 =& - \frac{\lambda_1}{2}\, U^2
- \left( \frac{\lambda_2}{2} + 2 \lambda_5 \right) T^2
\no &
- \left( \lambda_3 + \lambda_4 + \lambda_8 \right) U T
+ \Lambda \sqrt{U T}.
\end{align}
\es
Solution~\eqref{stab0} is the vacuum if
\be
\Lambda < 0 \quad \mbox{and} \quad \Lambda < - 4 \lambda_8 \sqrt{U T}.
\ee
Solution~\eqref{stabpi} has the lowest $V_0$ if
\be
\Lambda > 0 \quad \mbox{and} \quad \Lambda > 4 \lambda_8 \sqrt{U T}.
\ee
The vacua~\eqref{stab0} and~\eqref{stabpi} are related between themselves
through a change of the signs of both $\lambda_6$ and $\lambda_7$
(and hence of $\Lambda$ in Eq.~\eqref{Lambda}).

Either Eqs.~\eqref{stab0} or Eqs.~\eqref{stabpi} do not determine $U$ and $T$
uniquely as functions of the parameters of the potential;
this is investigated in Appendix~\ref{uniq}.

\section{The masses of the scalars}
\label{sec:scalarmasses}

\subsection{The mass of the double-charged scalar}

The double-charged scalar field is $c$.
Its mass terms are given by
\ba
V &=& \cdots + C \left[ \mu_2^2 + \left( \lambda_2 + 6 \lambda_5 \right) T
  + \left( \lambda_3 - 3 \lambda_4 \right) U \right.
\no &&
 \left. - 3 \lambda_7 \sqrt{U T} \cos{\theta} \right].
\ea
Let
$\mu_{++}$ be the mass of the double-charged scalar.
We have
\bs
\begin{align}
\mu_{++}^2 =& 
\left( - 4 \lambda_4 + \lambda_8
+ \frac{\lambda_6 \lambda_7}{\lambda_8} \right) U
+ \left( 2 \lambda_5 + \frac{\lambda_7^2}{\lambda_8} \right) T, \hspace*{-1mm}
\label{87a} \\
\mu_{++}^2 =&  \left( - 4 \lambda_4 - \lambda_8 \right) U
+ 2 \lambda_5 T
- \frac{\lambda_6 U}{2T}\, \sqrt{U T}
\no &
- \frac{9 \lambda_7}{2}\, \sqrt{U T},
\label{87b}
\end{align}
\es
for solutions~\eqref{stab} and~\eqref{stab0},
respectively.
For solution~\eqref{stabpi} one just has to use Eq.~\eqref{87b}
with inverted signs of $\lambda_6$ and $\lambda_7$.

\subsection{The masses of the single-charged scalars}

The scalar fields with electric charge $+1$ are $a$,
$d$,
and $f^*$.

Remember Eqs.~\eqref{UTtheta}.
We perform a gauge transformation to set $t$ real and positive.
Then,
\be
\left\langle 0 \left| b \right| 0 \right\rangle
= \exp (i \theta) \sqrt{U},
\quad
\left\langle 0 \left| e \right| 0 \right\rangle = \sqrt{T}.
\ee
We moreover define the field $\widehat a$ through
\be
a = \exp (i \theta)\, \widehat{a}.
\ee
The charged Goldstone boson then is
\be
P_1^+ = \frac{\sqrt{U}\, \widehat{a} + 2 \sqrt{T}\, d
  - \sqrt{3 T}\, f^\ast}{v_\mathrm{SM}},
\ee
where the normalization factor $v_\mathrm{SM}$
has been defined in Eq.~\eqref{vSM}.
We additionally define
\bs
\label{93kkr}
\ba
P_2^+ &\equiv& \frac{\sqrt{3}\, d + 2 f^\ast}{\sqrt{7}},
\\
P_3^+ &\equiv& \frac{7 \sqrt{T}\ \widehat{a} - 2 \sqrt{U}\, d
  + \sqrt{3 U}\, f^\ast}{\sqrt{7}\, v_\mathrm{SM}},
\ea
\es
which form,
together with $P_1^+$,
an orthonormal basis.
The mass terms of the single-charged scalars are then necessarily of the form
\be
V = \cdots + \left( \begin{array}{cc} P_2^-, & \hspace*{-1mm} P_3^- \end{array} \right)
\hspace*{-1mm}\left( \begin{array}{cc} \mathcal{M}_{11} & \hspace*{-2mm} \mathcal{M}_{12} + i c_x \\
  \mathcal{M}_{12} - i c_x & \hspace*{-2mm} \mathcal{M}_{22} \end{array} \right)
\hspace*{-1mm}\left( \begin{array}{c} P_2^+ \\ P_3^+ \end{array} \right).
\label{vnifofo}
\ee
The imaginary part $c_x$ only exists
in the case of spontaneous $CP$ breaking,
\textit{i.e.}\ in the case~\eqref{stab}.
In that case,
\be
c_x \equiv \frac{v_\mathrm{SM}\, \widetilde \Lambda}{4}\,
\sqrt{\frac{3}{T}}\, \sin{\theta},
\ee
where
\be
\label{tildeLambda}
\widetilde \Lambda \equiv \lambda_6 U - \lambda_7 T.
\ee
In the $CP$-conserving solutions~\eqref{stab0} and~\eqref{stabpi},
$c_x$ is zero.

For solution~\eqref{stab} one obtains
\bs
\label{opa}
\ba
\mathcal{M}_{11} &=&
\frac{\left( 2 \lambda_4 + 13 \lambda_8 \right) U}{7} - 7 \lambda_5 T
\no &&
+ \frac{\Lambda \left( - 3 \lambda_6 U + 7 \lambda_7 T \right)}{28
  \lambda_8 T},
\label{opa1} \\
\mathcal{M}_{22} &=& v_\mathrm{SM}^2 \left(
\frac{\lambda_8 - 2 \lambda_4}{7}
+ \frac{3 \lambda_6 \Lambda}{28 \lambda_8 T} \right),
\label{opa2} \\
\mathcal{M}_{12} &=& \sqrt{3 U}\, v_\mathrm{SM} \left[
  \frac{16 \lambda_4 - \lambda_8}{14} \right.
\no &&
\phantom{WWWWl} \left. + \frac{\left( \lambda_6 U + 7 \lambda_7 T \right) \Lambda}{112
    \lambda_8  U T} \right].
\ea
\es

With the $CP$-conserving solution~\eqref{stab0} one has
\bs
\label{iv0f0f}
\ba
\mathcal{M}_{11} &=& \frac{2 \lambda_4 - 13 \lambda_8}{7}\, U
- 7 \lambda_5 T 
\no &&
+ \left( - \frac{\lambda_6 U}{2 T} - \frac{27 \lambda_7}{14}
\right) \sqrt{U T},
\\
\mathcal{M}_{22} &=& v_\mathrm{SM}^2 \left[
  - \frac{2 \lambda_4 + \lambda_8}{7} \right.
\no &&
\phantom{WW}  \left. - \left( \frac{\lambda_6}{2 T} + \frac{\lambda_7}{14 U} \right)
  \sqrt{U T} \right],
\\
\mathcal{M}_{12} &=& \frac{\sqrt{3}}{14}\, v_\mathrm{SM}
\left[ \left( 16 \lambda_4 + \lambda_8 \right) \sqrt{U}
  - 3 \lambda_7 \sqrt{T} \right]. \hspace*{3mm}
\ea
\es
In the other $CP$-conserving solution~\eqref{stabpi},
Eqs.~\eqref{iv0f0f} with $\lambda_6 \to - \lambda_6$
and $\lambda_7 \to - \lambda_7$ hold.

\subsection{The masses of the neutral scalars}

We expand
\be
\label{expand2}
b = \exp (i \theta) \left(
\sqrt{U} + \frac{b_R + i b_I}{\sqrt{2}} \right),
\quad
e = \sqrt{T} + \frac{e_R + i e_I}{\sqrt{2}},
\ee
where $b_R$,
$b_I$,
$e_R$,
and $e_I$ are real scalar fields.
We define
\be
\label{Gzero}
\left( \begin{array}{c} G^0 \\ P \end{array} \right)
= \frac{1}{\sqrt{U + T}}
\left( \begin{array}{cc} \sqrt{U} & \sqrt{T} \\ - \sqrt{T} & \sqrt{U}
\end{array} \right)
\left( \begin{array}{c} b_I \\ e_I \end{array} \right).
\ee
The field $G^0$ is the neutral Goldstone boson;
$P$ is orthogonal to it.

\subsubsection{The CP-conserving cases}

In the $CP$-conserving cases~\eqref{stab0} and~\eqref{stabpi},
$P$ is a pseudoscalar while $b_R$ and $e_R$ are scalars.
The mass $m_P$ of $P$ is given by
\be
m_P^2 = - 2 \left( \lambda_8 + \frac{\Lambda}{4 \sqrt{U T}} \right)
\left( U + T \right)
\label{nfpppd}
\ee
in the case~\eqref{stab0};
in the case~\eqref{stabpi},
$m_P^2$ is given by Eq.~\eqref{nfpppd} with $\Lambda \to - \Lambda$.
Notice that the $m_P^2$ of Eq.~\eqref{nfpppd}
becomes zero when $\Lambda = - 4 \lambda_8 \sqrt{U T}$,
\textit{i.e.}\ at the point where the $CP$-conserving solution~\eqref{stab0}
gives way to the $CP$-breaking solution~\eqref{stab}.

The scalars $b_R$ and $e_R$ mix according to
\be
\left( \begin{array}{c} b_R \\ e_R \end{array} \right)
=
\left( \begin{array}{cc} \cos{\zeta} & - \sin{\zeta} \\
  \sin{\zeta} & \cos{\zeta} \end{array} \right)
\left( \begin{array}{c} h \\ H \end{array} \right),
\label{beta}
\ee
where $h$ and $H$ are two physical scalars with squared masses $M_h$ and $M_H$,
respectively,
and $\zeta$ is a mixing angle.
Clearly,
\ba
V
&=& \cdots + \frac{1}{2}
\left( \begin{array}{cc} b_R, & e_R \end{array} \right)
\bar M \left( \begin{array}{c} b_R \\ e_R \end{array} \right) \no
&=& \cdots + \frac{1}{2} \left( M_h h^2 + M_H H^2 \right),
\ea
where $\bar M$ is a $2 \times 2$ real symmetric matrix.
Therefore,
\bs
\label{asereje}
\ba
\bar M_{11} &=& M_h \cos^2{\zeta} + M_H \sin^2{\zeta}, \\
\bar M_{22} &=& M_h \sin^2{\zeta} + M_H \cos^2{\zeta}, \\
\bar M_{12} &=& \left( M_h - M_H \right) \cos{\zeta} \sin{\zeta}.
\ea
\es
It follows that
\bs
\label{wq}
\ba
\mathrm{tr}\, \bar M &=& \frac{M_h \left( \cos^2{\zeta} - \sin^2{\zeta} \right)
  + \bar M_{22}}{\cos^2{\zeta}},
\label{wq2} \\
M_H &=& \frac{\bar M_{22} - M_h \sin^2{\zeta}}{\cos^2{\zeta}},
\label{wq3} \\
\bar M_{12} &=& \left( M_h - \bar M_{22} \right) \tan{\zeta}.
\label{wq4}
\ea
\es
Using Eqs.~\eqref{stab0} one obtains
\bs
\label{0cfjvv}
\ba
\bar M_{11} &=& 2 \lambda_1 U
+ \left( 3 \lambda_6 - \frac{T}{U}\, \lambda_7 \right) \frac{\sqrt{U T}}{2},
\\
\bar M_{22} &=& 2 \left( \lambda_2 + 4 \lambda_5 \right) T
+ \left( 3 \lambda_7 - \frac{U}{T}\, \lambda_6 \right) \frac{\sqrt{U T}}{2}, \hspace*{7mm}
\label{M22oo} \\
\bar M_{12} &=& 2 \left( \lambda_3 + \lambda_4 + \lambda_8 \right) \sqrt{U T}
+ \frac{3}{2}\, \Lambda.
\label{M12oo}\ea
\es
With the solution~\eqref{stabpi},
Eqs.~\eqref{0cfjvv} with $\lambda_6 \to - \lambda_6$,
$\lambda_7 \to - \lambda_7$,
and $\Lambda \to - \Lambda$ hold.

\subsubsection{The $CP$-breaking case}

In the case~\eqref{stab} of spontaneous $CP$ violation,
$P$ mixes with $b_R$ and $e_R$:
\be
\label{matrixO}
\left( \begin{array}{c} b_R \\ e_R \\ P \end{array} \right)
= O
\left( \begin{array}{c} h \\ H \\ \mathcal{H} \end{array} \right),
\ee
where $h$,
$H$,
and $\mathcal{H}$ are three physical scalars with squared masses $M_h$,
$M_H$,
and $M_\mathcal{H}$,
respectively,
and $O$ is a real orthogonal $3 \times 3$ matrix.
One has
\ba
V
&=& \cdots + \frac{1}{2}
\left( \begin{array}{ccc} b_R, & e_R, & P \end{array} \right)
\tilde M \left( \begin{array}{c} b_R \\ e_R \\ P \end{array} \right) \no
&=& \cdots + \frac{1}{2} \left( M_h h^2 + M_H H^2 + M_\mathcal{H} \mathcal{H}^2
\right),
\ea
where $\tilde M$ is a $3 \times 3$ real symmetric matrix given by
\bs
\label{eq:u}
\ba
\tilde M_{11} &=&
2 \lambda_1 U
+ \frac{\lambda_7 T - 3 \lambda_6 U}{8 \lambda_8 U}\, \Lambda,
\label{u11} \\
\tilde M_{22} &=& 2 \left( \lambda_2 + 4 \lambda_5 \right) T
+ \frac{\lambda_6 U - 3 \lambda_7 T}{8 \lambda_8 T}\, \Lambda,
\label{u22} \\
\tilde M_{12} &=&
2 \left( \lambda_3 + \lambda_4 - \lambda_8 \right) \sqrt{U T}
- \frac{\Lambda^2}{8 \lambda_8 \sqrt{U T}},
\label{u12} \\
\tilde M_{33} &=& 2 \left( U + T \right)
\left( \lambda_8 - \frac{\Lambda^2}{16 \lambda_8 U T} \right),
\label{u33} \\
\tilde M_{13} &=& \frac{\sqrt{U + T}}{2 \sqrt{U}}\
\widetilde \Lambda\, \sin{\theta},
\label{u13} \\
\tilde M_{23} &=& - \frac{\sqrt{U + T}}{2 \sqrt{T}}\
\widetilde \Lambda\, \sin{\theta}.
\label{u23}
\ea
\es
Notice that in the $CP$-conserving limit $\left| \cos{\theta} \right| = 1$
one has $\tilde M_{13} = \tilde M_{23} = \tilde M_{33} = 0$.
In that limit $P$ becomes a massless physical pseudoscalar,
just as was pointed out after Eq.~\eqref{nfpppd}.

Since
\be
O^T \tilde M O = \mathrm{diag} \left( M_h, M_H, M_\mathcal{H} \right),
\label{otmo}
\ee
one has
\bs
\label{kdoooe}
\ba
O_{11}^2 + O_{12}^2 + O_{13}^2 &=& 1, \\
M_h O_{11}^2 + M_H O_{12}^2 + M_\mathcal{H} O_{13}^2 &=& \tilde M_{11}, \\
M_h^2 O_{11}^2 + M_H^2 O_{12}^2 + M_\mathcal{H}^2 O_{13}^2 &=&
\left( \tilde M^2 \right)_{11}.
\ea
\es
Solving the set~\eqref{kdoooe} of three linear equations,
one obtains
\be
O_{11}^2 = \frac{M_h \left( \tilde M^2 \right)_{11} +
  M_h \left( M_h - \mathrm{tr}\, \tilde M \right) \tilde M_{11}
  + \det \tilde M}{M_h^2\
  \mathrm{tr}\, \tilde M - 2 M_h\, I_2 \left( \tilde M \right)
  + 3 \det \tilde M},
\label{o11}
\ee
where
\ba
I_2 \left( \tilde M \right) &=&
\tilde M_{11} \tilde M_{22}
+ \tilde M_{11} \tilde M_{33}
+ \tilde M_{22} \tilde M_{33}
\no &&
- \left( \tilde M_{12} \right)^2
- \left( \tilde M_{13} \right)^2
- \left( \tilde M_{23} \right)^2.
\ea
Equation~\eqref{o11} may be rewritten
\be
\hat a \tilde M_{11} + \hat b \tilde M_{12}
+ \hat c \left( \tilde M_{12} \right)^2 = \hat d,
\label{q1}
\ee
where
\bs
\ba
\hat a &=& M_h^2
- M_h \left( \tilde M_{22} + \tilde M_{33} \right)
+ \tilde M_{22} \tilde M_{33} - \left( \tilde M_{23} \right)^2
\no & &
- \, O_{11}^2 \left\{ M_h^2 - 2 M_h \left( \tilde M_{22} + \tilde M_{33} \right) \right.
\no && \phantom{WWW}
+ \left. 3 \left[ \tilde M_{22} \tilde M_{33} - \left( \tilde M_{23} \right)^2
\right] \right\},
\\
\hat b &=& 2 \tilde M_{13} \tilde M_{23} \left( 1 - 3 O_{11}^2 \right),
\\
\hat c &=& M_h - \tilde M_{33} - O_{11}^2 \left( 2 M_h - 3 \tilde M_{33} \right),
\\
\hat d &=&
\left( \tilde M_{22} - M_h \right) \left( \tilde M_{13} \right)^2
+ O_{11}^2 M_h \left\{ M_h \left( \tilde M_{22} + \tilde M_{33} \right) \right.
\no &&
\left. + 2 \left[ \left( \tilde M_{13} \right)^2 + \left( \tilde M_{23} \right)^2
  - \tilde M_{22} \tilde M_{33} \right] \right\}.
\hspace*{7mm}
\ea
\es
One also knows that $M_h$ is a root of the eigenvalue equation of $\tilde M$,
\textit{viz.}\ that
%
\be
M_h^3 - M_h^2\ \mathrm{tr}\, \tilde M
+ M_h\, I_2 \left( \tilde M \right) - \det \tilde M = 0.
\label{mjkg}
\ee
Equation~\eqref{mjkg} may be rewritten
\be
\hat a' \tilde M_{11} + \hat b' \tilde M_{12}
+ \hat c' \left( \tilde M_{12} \right)^2 = \hat d',
\label{q2}
\ee
where
\bs
\ba
\hat a' &=& \left( \tilde M_{23} \right)^2
- \tilde M_{22} \tilde M_{33} + M_h \left( \tilde M_{22} + \tilde M_{33} \right)
\no &&
- M_h^2,
\\
\hat b' &=& - 2 \tilde M_{13} \tilde M_{23},
\\
\hat c' &=& \tilde M_{33} - M_h,
\\
\hat d' &=& - M_h^3
+ M_h^2 \left( \tilde M_{22} + \tilde M_{33} \right)
- \tilde M_{22} \left( \tilde M_{13} \right)^2
\no & &
+ M_h \left[ \left( \tilde M_{13} \right)^2 + \left( \tilde M_{23} \right)^2
  - \tilde M_{22} \tilde M_{33} \right].
\ea
\es
Equations~\eqref{q1} and~\eqref{q2} together 
produce a quadratic equation for $\tilde M_{12}$:
\be
\left( \hat a \hat c' - \hat c \hat a' \right) \left( \tilde M_{12} \right)^2
+ \left( \hat a \hat b' - \hat b \hat a' \right) \tilde M_{12}
+ \left( \hat d \hat a' - \hat a \hat d' \right) = 0.
\label{q4}
\ee
This allows one to determine $\tilde M_{12}$.
Afterwards,
\be
\tilde M_{11} = \frac{\left( \hat b \hat c' - \hat c \hat b' \right)
  \tilde M_{12}
  + \hat c \hat d' - \hat d \hat c'}{\hat c \hat a' - \hat a \hat c'}
\label{q5}
\ee
fixes $\tilde M_{11}$.

\section{The three- and four-Higgs vertices}
\label{sec:g3g4}

We plug Eqs.~\eqref{expand2} into Eq.~\eqref{vacpot},
keeping only the terms that are either three-linear or four-linear
in the fields $b_R$,
$b_I$,
$e_R$,
and $e_I$.
We find the tri-linear terms
\begin{widetext}
\bs
\label{vtres}
\ba
\left[ V_{b,e} \right]_{(3)} &=& \frac{1}{2 \sqrt{2}} \left\{
2 \lambda_1 \sqrt{U}\, b_R \left( b_R^2 + b_I^2 \right) 
+ 2 \left( \lambda_2 + 4 \lambda_5 \right) \sqrt{T}\,
e_R \left( e_R^2 + e_I^2 \right)
\right. 
\\ & &
+ 2 \left( \lambda_3 + \lambda_4 \right) \left[
  \sqrt{U}\, b_R \left( e_R^2 + e_I^2 \right)
  + \sqrt{T}\, e_R \left( b_R^2 + b_I^2 \right)
  \right]
\\ & &
+ \lambda_6 \cos{\theta} \left[
  \sqrt{U} \left( 3 b_R^2 e_R + 2 b_R b_I e_I + b_I^2 e_R \right)
  + \sqrt{T}\, b_R \left( b_R^2 + b_I^2 \right) \right]
\\ & &
+ \lambda_7 \cos{\theta} \left[
  \sqrt{U}\, e_R \left( e_R^2 + e_I^2 \right)
  + \sqrt{T} \left( 3 b_R e_R^2 + 2 b_I e_R e_I + b_R e_I^2 \right) \right]
\\ & &
+ 2 \lambda_8 \cos{\left( 2 \theta \right)} \left[
  \sqrt{U} \left( b_R e_R^2 - b_R e_I^2 + 2 b_I e_R e_I \right)
  + \sqrt{T} \left( b_R^2 e_R - b_I^2 e_R + 2 b_R b_I e_I \right) \right]
\\ & &
+ \lambda_6 \sin{\theta} \left[
  \sqrt{U} \left( 3 b_R^2 e_I - 2 b_R b_I e_R + b_I^2 e_I \right)
  - \sqrt{T}\, b_I \left( b_R^2 + b_I^2 \right) \right]
\\ & &
+ \lambda_7 \sin{\theta} \left[
  \sqrt{U}\, e_I \left( e_R^2 + e_I^2 \right)
  + \sqrt{T} \left( - 3 b_I e_R^2 + 2 b_R e_R e_I - b_I e_I^2 \right) \right]
\\ & & \left.
+ 4 \lambda_8 \cos{\theta} \sin{\theta} \left[
  \sqrt{U} \left( b_I e_I^2 - b_I e_R^2 + 2 b_R e_R e_I \right)  
  + \sqrt{T} \left( b_R^2 e_I - b_I^2 e_I - 2 b_R b_I e_R \right) \right]
\right\}
\\ &=& g_3 h^3 + \cdots
\ea
\es
%
and the four-linear terms
%
\bs
\label{vquatro}
\ba
\left[ V_{b,e} \right]_{(4)} &=& \frac{1}{8} \left\{
\lambda_1 \left( b_R^2 + b_I^2 \right)^2
+ \left( \lambda_2 + 4 \lambda_5 \right) \left( e_R^2 + e_I^2 \right)^2
\right. 
+ 2 \left( \lambda_3 + \lambda_4 \right)
\left( b_R^2 + b_I^2 \right) \left( e_R^2 + e_I^2 \right)
\\ & &
+ 2 \lambda_6 \cos{\theta}
\left( b_R e_R + b_I e_I \right) \left( b_R^2 + b_I^2 \right)
+ 2 \lambda_7 \cos{\theta}
\left( b_R e_R + b_I e_I \right) \left( e_R^2 + e_I^2 \right)
\\ & &
+ 2 \lambda_8 \cos{\left( 2 \theta \right)}
\left[ \left( b_R^2 - b_I^2 \right) \left( e_R^2 - e_I^2 \right) + 4 b_R b_I e_R e_I \right]
\\ & & \left.
+ 2 \left[ \lambda_6 \left( b_R^2 + b_I^2 \right)
  + \lambda_7 \left( e_R^2 + e_I^2 \right)
  + 4 \lambda_8 \cos{\theta} \left( b_R e_R + b_I e_I \right)
  \vphantom{\lambda_6 \left( b_R^2 + b_I^2 \right)} \right]
\left( b_R e_I - b_I e_R \right) \sin{\theta}
\vphantom{\left( b_R^2 + b_I^2 \right)^2} \right\}
\\ &=& g_4 h^4 + \cdots,
\ea
\es
\end{widetext}
respectively.

In the cases $\theta = 0$ and $\theta = \pi$ one has,
according to Eq.~\eqref{beta},
\bs
\ba
b_R &=& h \cos{\zeta} + \cdots,
\\
e_R &=& h \sin{\zeta} + \cdots,
\ea
\es
while $b_I$ and $e_I$ have no $h$ component.
Therefore,
if $\theta = 0$ then
\bs
\label{g3(1)}
\ba
g_3 &=& \frac{1}{2 \sqrt{2}}
\left[
  2 \lambda_1 \sqrt{U} \cos^3{\zeta}
  \right. \\ & &
  + 2 \left( \lambda_2 + 4 \lambda_5 \right) \sqrt{T} \sin^3{\zeta}
  \\ & &
  + 2 \left( \lambda_3 + \lambda_4 + \lambda_8 \right) \sin{\zeta} \cos{\zeta}
  \\ & &
  \times \left( \sqrt{U} \sin{\zeta} + \sqrt{T} \cos{\zeta} \right)
  \\ & &
  + \lambda_6 \cos^2{\zeta}
  \left( 3 \sqrt{U} \sin{\zeta} + \sqrt{T} \cos{\zeta} \right)
  \\ & & \left.
  + \lambda_7 \sin^2{\zeta}
  \left( \sqrt{U} \sin{\zeta} + 3 \sqrt{T} \cos{\zeta} \right)
  \right], 
\ea
\es
and
\bs
\label{g4(1)}
\ba
g_4 &=& \frac{1}{8} \left[
  \lambda_1 \cos^4{\zeta}
  + \left( \lambda_2 + 4 \lambda_5 \right) \sin^4{\zeta}
  \right. \\ & &
  + 2 \left( \lambda_3 + \lambda_4 + \lambda_8 \right) \sin^2{\zeta} \cos^2{\zeta}
  \\ & & \left.
  + 2 \left( \lambda_6 \cos^2{\zeta} + \lambda_7 \sin^2{\zeta} \right)
  \sin{\zeta} \cos{\zeta} \right].
\ea
\es
If $\theta = \pi$,
then Eqs.~\eqref{g3(1)} and~\eqref{g4(1)} with $\lambda_6 \to - \lambda_6$
and $\lambda_7 \to - \lambda_7$ hold.

In the $CP$-violating case one must,
according to Eqs.~\eqref{Gzero} and~\eqref{matrixO},
effect the transformations
\bs
\label{120}
\ba
b_R &\to& O_{11} h, \\
e_R &\to& O_{21} h, \\
b_I &\to& - \sqrt{\frac{T}{U+T}}\ O_{31} h, \\
e_I &\to& \sqrt{\frac{U}{U+T}}\ O_{31} h
\ea
\es
in Eqs.~\eqref{vtres} and~\eqref{vquatro}
to obtain the values of $g_3$ and $g_4$,
respectively.
Of course,
one must moreover employ Eqs.~\eqref{stab}.

\section{Procedure}
\label{sec:procedure}

In our practical procedure we follow the following steps:
\begin{enumerate}
\item We input $\alpha_\mathrm{em}$,
  $G_F$,
  $m_Z$,
  and $m_W$.
  In our numerical analysis we have used Eqs.~\eqref{theinputs}
  for the first three quantities,
  while we have let $m_W$ vary between its SM value
  $m_W^\mathrm{SM} = 79.829\,\mathrm{GeV}$
  and a maximum value $80.5\,\mathrm{GeV}$.
\item We compute $U$,
  $T$,
  and $v_\mathrm{SM}$ by using Eqs.~\eqref{3} and~\eqref{vSM},
  respectively.
\item We input $\lambda_2$,
  $\lambda_4$,
  $\lambda_5$,
  $\lambda_6$,
  $\lambda_7$,
  $\lambda_8$,
  and $M_h = \left( 125\,\mathrm{GeV} \right)^2$.
\item We determine $\Lambda$ and $\widetilde \Lambda$
  by using Eqs.~\eqref{Lambda} and~\eqref{tildeLambda},
  respectively.
\item The scalar $h$ couples to  gauge-boson pairs through
  \ba
  \mathcal{L} &=& \cdots + \sqrt{\frac{4 \pi \alpha_\mathrm{em}}{1 - c_W^2}}\ h
  \biggl( \kappa_W m_W W_\mu^+ W^{\mu -}
  \no && \left.
  + \kappa_Z\, \frac{m_Z}{2 c_W}\, Z_\mu Z^\mu \right),
  \ea
  which defines the coupling modifiers\footnote{They are called
  `modifiers' because in the SM $\kappa_W = \kappa_Z = 1$.}
  $\kappa_W$ and $\kappa_Z$.
  Those experimentally measured quantities are,
  in our model,
  equal to $\cos{\zeta}$.
  In our computations we enforce the condition
  \be
  - \arccos{0.9} < \zeta < \arccos{0.9},
  \ee
  that approximately reproduces the experimental findings~\cite{CMS:2022dwd};
  in some cases,
  we use the more stringent condition~\cite{CMS:2022dwd}
  \be
  0.97 \le \cos \zeta \le 1.
  \label{kappasWZ}
  \ee
\item We determine which regime we are in.
  If
  \be
  \lambda_8 > 0 \quad \mbox{and} \quad
  - 4 \lambda_8 \sqrt{U T} < \Lambda < 4 \lambda_8 \sqrt{U T}, \hspace*{-5mm}
  \ee
  then we are in the $CP$-violating regime of Eqs.~\eqref{stab}.
  If
  \be
  \Lambda < - 4 \lambda_8 \sqrt{U T} \quad \mbox{and} \quad \Lambda < 0,
  \ee
  then we are in the $CP$-conserving regime of Eqs.~\eqref{stab0}.
  If
  \be
  \Lambda > 4 \lambda_8 \sqrt{U T} \quad \mbox{and} \quad \Lambda > 0,
  \ee
  then we are in the $CP$-conserving regime of Eqs.~\eqref{stabpi}.
\end{enumerate}

\subsection{$\theta = 0$}
\label{subsec0}

If we are in the regime of Eqs.~\eqref{stab0},
then we proceed as follows.
We determine $\mu_{++}^2$ through Eq.~\eqref{87b}.
We determine $m_P^2$ through Eq.~\eqref{nfpppd}.
We determine the
(real and symmetric)
mass matrix $\mathcal{M}$ of the two charged scalars
through Eqs.~\eqref{iv0f0f};
we find the eigenvalues of that matrix,
\textit{viz.}\ the squared masses of the two charged scalars.
We compute $\bar M_{22}$ through Eq.~\eqref{M22oo}.
We use Eqs.~\eqref{wq} to determine $\mathrm{tr}\, \bar M$,
$M_H$ (which must turn out positive and sufficiently large),
and $\bar M_{12}$.
We determine $\lambda_1$ from Eqs.~\eqref{asereje}--\eqref{0cfjvv},
\textit{viz.}
\ba
\lambda_1 &=& \frac{1}{2 U} \left[
  \frac{M_h \left( \cos^2{\zeta} - \sin^2{\zeta} \right)
    + \bar M_{22} \sin^2{\zeta}}{\cos^2{\zeta}}
  \right. \no && \left.
  + \left( \frac{T}{U}\, \lambda_7 - 3 \lambda_6 \right) \frac{\sqrt{U T}}{2}
  \right].
\ea
We determine $\lambda_3$ from $\bar M_{12}$ as given by Eq.~\eqref{M12oo}.

We determine $\mu_1^2$,
$\mu_2^2$,
and $V_0$ from Eqs.~\eqref{stab0}.
We enforce both the condition $V_0 < 0$ and the conditions~\eqref{alternative};
if those conditions are not obeyed,
then the input is bad and must be discarded.

We then follow the procedure
in subsection~\ref{subsec:Alternative} of Appendix~\ref{uniq}.
We solve Eq.~\eqref{cons} for $U$;
one of the solutions must be the initial $U$,
but there may be other real and positive solutions.
Once one such solution has been found,
we extract $T$ from Eq.~\eqref{t2};
it should be positive too.
Next we compute
\be
G \equiv \frac{2 U \left[ \mu_1^2 + \lambda_1 U
    + \left( \lambda_3 + \lambda_4 + \lambda_8 \right)
    T \right]}{- 3 \lambda_6 U - \lambda_7 T},
\ee
We check that $G^2 = U T$ and that
\ba
\mu_2^2 &=& - \left( \lambda_2 + 4 \lambda_5 \right) T
- \left( \lambda_3 + \lambda_4 + \lambda_8 \right) U
\no &&
- \left( \frac{\lambda_6 U}{2 T}
+ \frac{3 \lambda_7}{2} \right) G.
\ea
Finally,
we compute
\ba
\tilde V_0 &=& - \frac{\lambda_1}{2}\, U^2
- \left( \frac{\lambda_2}{2} + 2 \lambda_5 \right) T^2
- \left( \lambda_3 + \lambda_4 + \lambda_8 \right) U T
\no &&
- \left( \lambda_6 U + \lambda_7 T \right) G.
\ea
We check whether $\tilde V_0 > V_0$;
if this does not hold,
then the input is bad and must be discarded.

Last but not least,
we use Eqs.~\eqref{g3(1)} and~\eqref{g4(1)} to find out the values of $g_3$ and $g_4$.

\subsection{$\theta = \pi$}

If we are in the regime of Eqs.~\eqref{stabpi},
then we follow exactly the same procedure as in subsection~\ref{subsec0},
except that in all the equations employed we make $\lambda_6 \to - \lambda_6$,
$\lambda_7 \to - \lambda_7$,
and $\Lambda \to - \Lambda$.

\subsection{$CP$-violating regime}

If we are in the regime of Eqs.~\eqref{stab},
then we proceed as follows.
We determine $\mu_{++}^2$ through Eq.~\eqref{87a}.
We determine the
(complex and Hermitian)
mass matrix $\mathcal{M}$ of the two charged scalars
through Eqs.~\eqref{vnifofo}--\eqref{opa};
we find the eigenvalues of that matrix,
\textit{viz.}\ the squared masses of the two charged scalars.
We determine $\tilde M_{22}$,
$\tilde M_{33}$,
$\tilde M_{13}$,
and $\tilde M_{23}$ through Eqs.~\eqref{u22},
\eqref{u33},
\eqref{u13},
and~\eqref{u23},
respectively.
Next we use Eqs.~\eqref{q1}--\eqref{q5} to determine firstly
$\tilde M_{12}$---by solving Eq.~\eqref{q4},
with a twofold ambiguity---and secondly $\tilde M_{11}$;
in those equations,
we use $O_{11}^2 = \cos^2{\zeta}$.
We now know the whole matrix $\tilde M$;
we diagonalize it through Eq.~\eqref{otmo} to determine both $M_H$
and $M_\mathcal{H}$
(which should be sufficiently heavy)
and the matrix $O$.

We determine $\lambda_3$ from $\tilde M_{12}$ by using Eq.~\eqref{u12},
and $\lambda_1$ from $\tilde M_{11}$ by using Eq.~\eqref{u11}.
We determine $\mu_1^2$,
$\mu_2^2$,
and $V_0$ from Eqs.~\eqref{stab}.
We enforce both $V_0 < 0$ and the conditions~\eqref{alternative}.
If those conditions are not obeyed,
then the input is bad and must be discarded.

In order to determine $g_3$ and $g_4$ we operate the transformation~\eqref{120}
on Eqs.~\eqref{vtres} and~\eqref{vquatro},
respectively.

\subsection{Vacuum stability}
\label{evade}
In addition to the UNI and BFB conditions,
and to the analytical vacuum-stability conditions
derived in Appendix~\ref{uniq}, 
in our numerical calculations we have also used
the package {\tt EVADE}~\cite{Hollik:2018wrr,Ferreira:2019iqb,EVADE}
to check that each of our points
is the global minimum of the corresponding SP. 
{\tt EVADE} identifies the tree-level minima
by using polynomial homotopy continuation,
employing {\tt HOM4PS2}~\cite{HOM4PS2}.

By requiring the electroweak (EW) vacuum to be the global minimum of the SP,
we are imposing a strong constraint on its parameters. 
That constraint may be relaxed if one just requires that
the lifetime of the EW vacuum
is sufficiently larger than the age of the Universe. 
In this case,
the EW vacuum is metastable but long-lived. 
If a minimum of the SP deeper than the EW vacuum is detected,
{\tt EVADE} evaluates the lifetime of the EW vacuum
based on the tunneling probability into the deeper minimum,
by using the straight-path approximation. 
That lifetime is quantified through the so-called `bounce action'.
In our analysis,
the parameter points are considered valid
only when the EW vacuum is the global minimum of the SP;
yet,
for comparison,
we also evaluate $g_3$ and $g_4$ for the cases
where the EW vacuum is sufficiently long-lived,
specifically when the bounce action is larger than 440~\cite{Hollik:2018wrr}.

\section{Results}
\label{sec:results}

We have performed a numerical analysis
for the three cases---\textit{viz.}\ Eqs.~\eqref{stab},
\eqref{stab0},
and~\eqref{stabpi}---mentioned in Section~\ref{sec:procedure}. 
That analysis has shown that the $CP$-violating case~\eqref{stab}
yields results that are difficult to reconcile
with the experimental constraints on the masses of scalars,
\textit{viz.}\ the lightest neutral scalar has a mass
that always turns out smaller than 30\,GeV\footnote{The reason for this is that,
in the limit where $\left| \cos{\theta} \right| \to 1$,
{\it i.e.}\ in the limit of no $CP$ violation,
one of the scalars becomes exactly massless,
\textit{cf.}\ the remarks after Eqs.~\eqref{nfpppd} and~\eqref{eq:u}.
In practice,
one never gets sufficiently far from that limit.}
and the lightest charged scalar has a mass
that is always less than 180\,GeV
(the other scalars can reach masses up to approximately
650\,GeV).\footnote{In the $CP$-violating case,
$\mu_2^2$ is given by Eq.~\eqref{69e},
which shows that it does not grow when $T \to 0$,
\textit{i.e.}\ when the inputted $m_W$ approaches $m_W^\mathrm{SM}$.
As a consequence of $\mu_2^2$ always remaining at the Fermi scale,
there is no way the scalar masses can become $\gtrsim$ 1\,TeV in that case.}
For this reason,
here \emph{we give results exclusively for the $CP$-conserving cases}
$\theta = 0$ and $\theta = \pi$.
Those cases differ only
in the signs of the parameters $\lambda_6$ and $\lambda_7$,
which do not affect any physical observable;
therefore,
we will treat them as \emph{a single $CP$-conserving case}.

\begin{figure}[t]
\begin{center}
\includegraphics[width=0.5\textwidth]{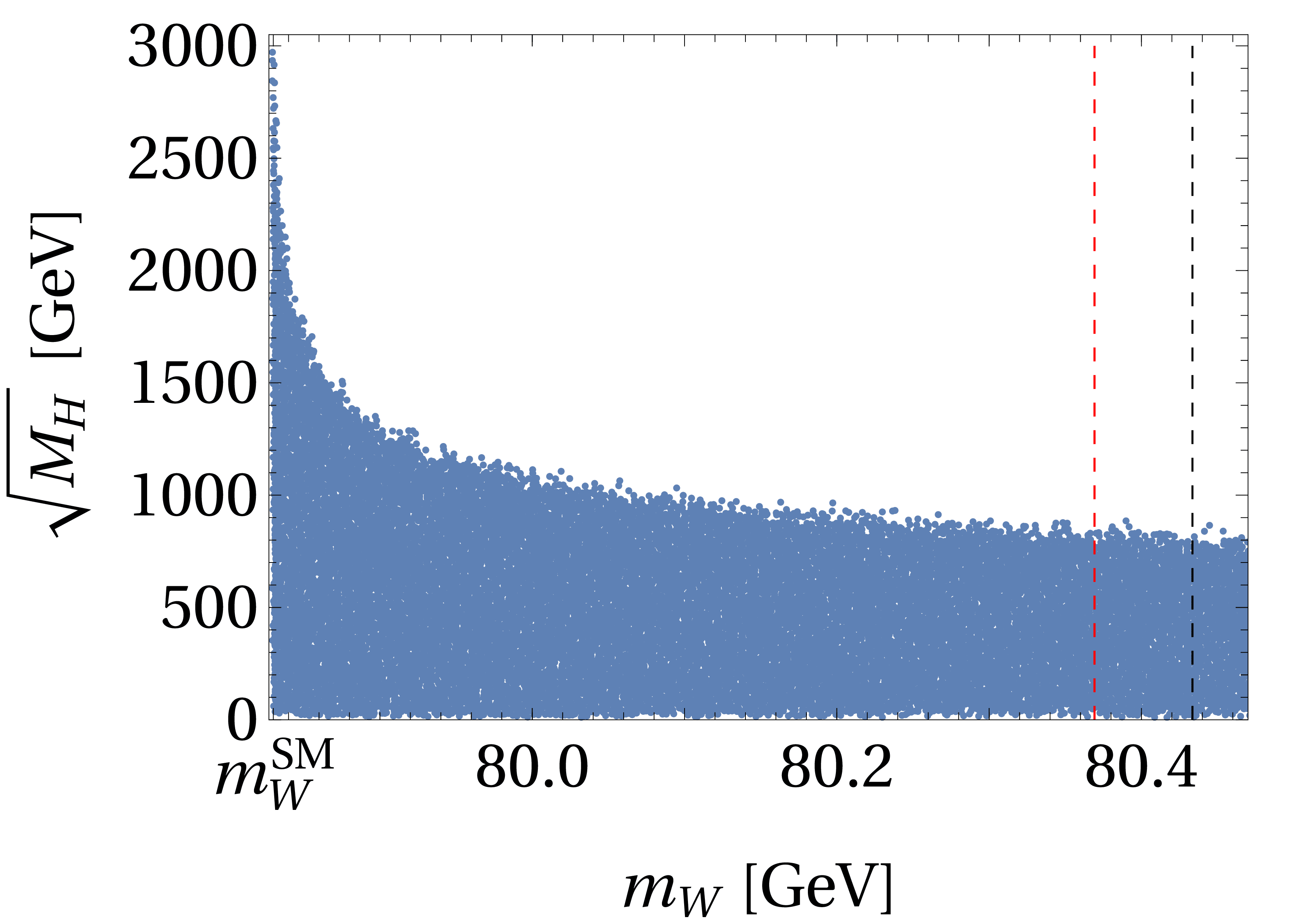}
\end{center}
\caption{Scatter plot of $\sqrt{M_H}$ \textit{versus}\/ $m_W$
  in the $CP$-conserving case.
  The red dashed line indicates the recent experimental world average
  of $m_W$~\cite{pdg};
  the blue dashed line indicates the $m_W$ value
  from the CDF measurements~\cite{CDF}.
  \label{fig:mH_mW}}
\end{figure}
In the equations for $\mu_2^2$ in the $CP$-conserving cases,
\textit{viz.}\ in Eqs.~\eqref{os2} and~\eqref{os4},
$T$ appears in the denominator of some terms.
Therefore,
in the limit $m_W \to m_W^\mathrm{SM}$,
\textit{i.e.}\ when $T$ approaches
zero---see Fig.~\ref{fig:VEVs_Y12}---$\mu_2^2$ becomes large
and the masses of the scalars can reach large values.
This is shown in Fig.~\ref{fig:mH_mW} through the behaviour of $\sqrt{M_H}$.

\begin{figure}[ht]
\begin{center}
\includegraphics[width=0.5\textwidth]{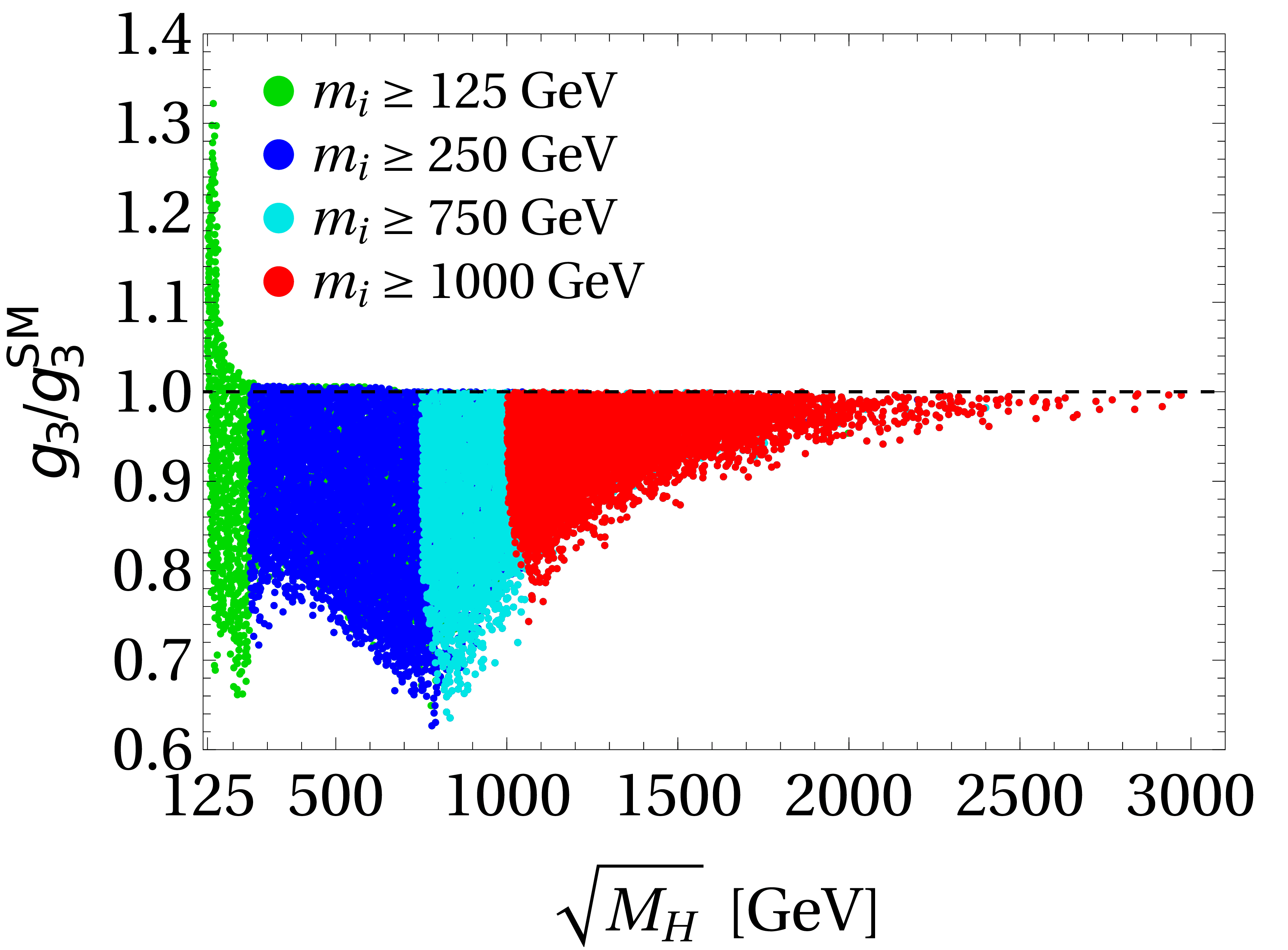}
\end{center}
\caption{Scatter plot of $g_3 \left/ g_3^\mathrm{SM} \right.$
  \textit{versus}\/ $\sqrt{M_H}$ for points with various lower bounds
  applied to all the scalar masses $m_i$ simultaneously.
  The dashed line marks $g_3 = g_3^\mathrm{SM}$. 
  \label{fig:g3_mH}}
\end{figure}

The masses of the scalars
impose strict constraints on the Higgs self-coupling $g_3$:
a lower bound on $g_3 \left/ g_3^\mathrm{SM} \right.$
exists for any value of the scalar masses.
The neutral scalar provides the strongest constraint,
effectively preventing $g_3 \left/ g_3^\mathrm{SM} \right.$
from being larger than 1 for $\sqrt{M_H} > 250$~GeV.
All this is shown in Fig.~\ref{fig:g3_mH}.
%

\begin{figure}[t]
\begin{center}
\includegraphics[width=0.5\textwidth]{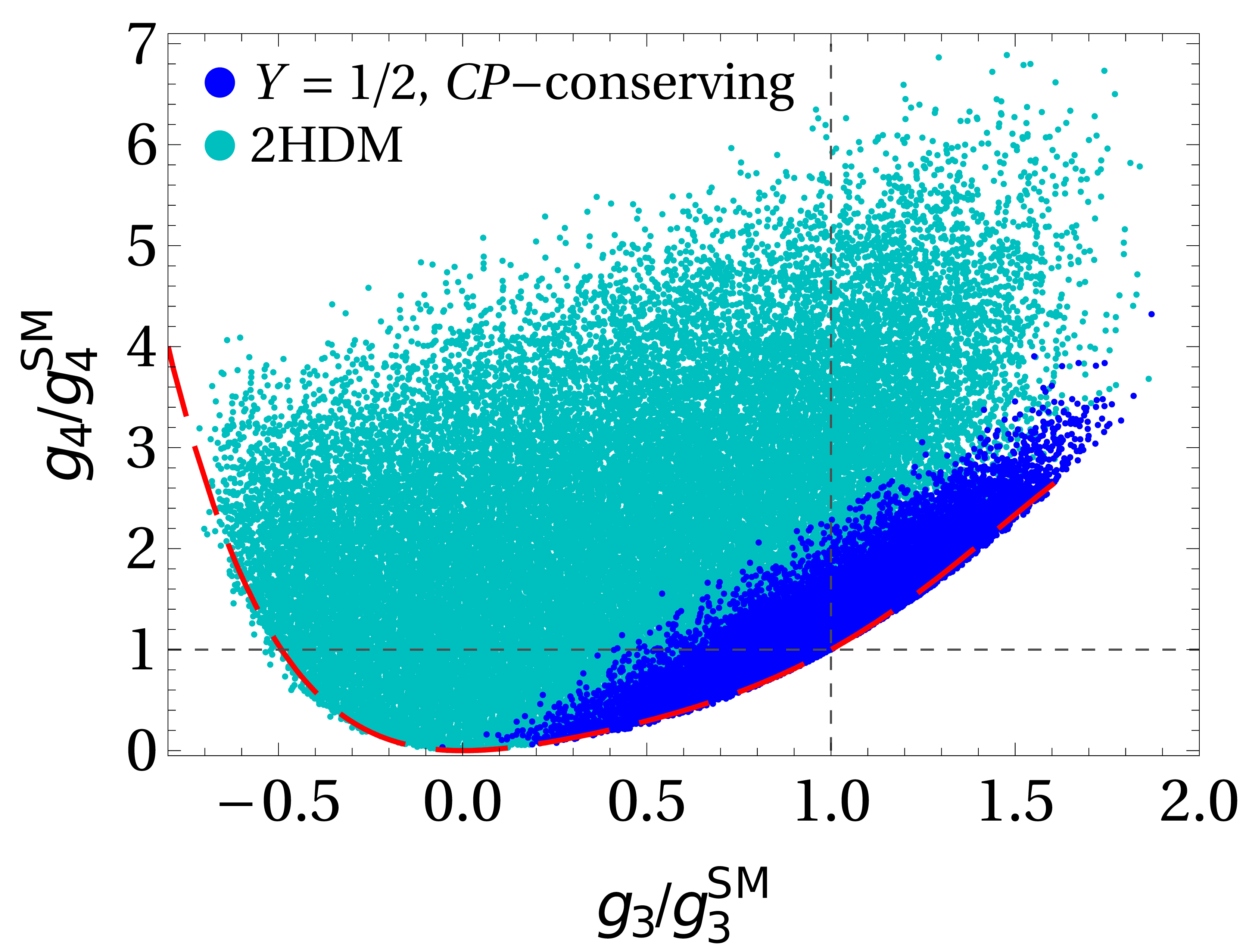}
\end{center}
\caption{Scatter plot of $g_4 \left/ g_4^\mathrm{SM} \right.$
  \textit{versus}\/ $g_3 \left/ g_3^\mathrm{SM} \right.$
  for the the 2HDM (cyan points)
  and the $CP$-conserving quadruplet model (blue points).
  In producing this figure we have utilized relaxed constraints,
  {\textit viz.}\ $\cos \zeta \ge 0.9$
  and all the masses of scalars larger than 10\,GeV.
  The straight dashed lines mark the SM values of $g_3$ and $g_4$.
  The red dashed line,
  given by Eq.~\eqref{g34_bound},
  marks the approximate boundary of the allowed region
  for $-0.6 < g_3 / g_3^\mathrm{SM} < 1.6$.
  Note that cyan points lie below the blue points
  and they also reach the red dashed line.
  \label{fig:g34_models}}
\end{figure}

A comparison of the predictions
for $g_4 \left/ g_4^\mathrm{SM} \right.$
and $g_3 \left/ g_3^\mathrm{SM} \right.$
in the two-Higgs-doublet model (2HDM)
and in the quadruplet model is shown in Fig.~\ref{fig:g34_models}. 
In producing that figure
we have required all the scalar masses to be larger than 10\,GeV;
furthermore,
the 2HDM points have been evaluated following Ref.~\cite{we}.
One observes that the quadruplet extension of the SM
allows much narrower ranges for $g_3$ and $g_4$ than the 2HDM. 
Interestingly,
the lower bound on $g_4 \left/ g_4^\mathrm{SM} \right.$
as a function of $g_3 \left/ g_3^\mathrm{SM} \right.$,
given in Ref.~\cite{we} by the (approximate) equation
\ba
\frac{g_4}{g_4^\mathrm{SM}}
&\ge& 2.07 \left( \frac{g_3}{g_3^\mathrm{SM}} \right)^2
  - 2.84 \left( \frac{g_3}{g_3^\mathrm{SM}} \right)^3
  \no &&
  + 2.44 \left( \frac{g_3}{g_3^\mathrm{SM}} \right)^4
  - 0.67 \left( \frac{g_3}{g_3^\mathrm{SM}} \right)^5,
\label{g34_bound}
\ea
also applies to the quadruplet model;
we display this lower bound in Fig.~\ref{fig:g34_models}
through a dashed red line.

In the left panel of Fig.~\ref{fig:g34_CP-cons}
a comparison of the allowed areas
for different bounds on $\cos{\zeta}$ is displayed. 
We observe that that parameter has a strong impact
on restricting $g_3$ and $g_4$.
As mentioned previously,
the scalar masses also have a strong impact on $g_3$ and $g_4$,
as displayed again in the right panel of Fig.~\ref{fig:g34_CP-cons}.

While the analytical stability conditions
derived in Appendix~\ref{uniq}
eliminate approximately 32.5\% of the otherwise allowed points,
the {\tt EVADE} test mentioned in subsection~\ref{evade}
additionally eliminates approximately $1.5\%$ of the points.
However,
about one third of the discarded points are classified by {\tt EVADE}
as long-lived points. 
If one allows those long-lived points,
then $g_3$ and $g_4$ can be up to 20\% larger,
though this percentage decreases for larger scalar masses.
So,
the distinction between stable and metastable points
seems not to be very important for this model.

\section{Discussion}
\label{sec:conclusions}

In this paper,
we have studied the scalar potential of the extension of the SM
through a scalar $SU(2)$ quadruplet with hypercharge either 1/2
(in the main body of the paper)
or 3/2 (in Appendix~\ref{Y32}). 
This extension of the SM has been studied previously
in Refs.~\cite{AbdusSalam:2013eya,Dawson:2017vgm},
with recent studies on scalar quadruplets with these specific hypercharges
in Refs.~\cite{roma1,roma2}.
Our results for $g_3$ are comparable to those of Ref.~\cite{kannikerecent}.

We have presented the full unitarity conditions.
Instead of attempting to derive fully analytical
bounded-from-below conditions,
we have suggested a method that involves a numerical scan of the phase space.
We wrote down some analytical vacuum-stability conditions,
which were complemented in numerical calculations
by using the {\tt EVADE} package. 
Analytical expressions
for the cubic and quartic Higgs couplings are also provided.

After numerical analysis, we have found the following:
\begin{itemize}
\item Only the $Y = 1/2$,
  $CP$-conserving case provides phenomenologically viable results. 
  The $Y = 1/2$,
  $CP$-violating case yields very low scalar masses,
  that are difficult to reconcile with phenomenological constraints,
  while the $Y = 3/2$ case contradicts experimental measurements of $m_W$.
\item The cubic and quartic Higgs self-couplings in the SM extension
  with a quadruplet are more restricted than in the 2HDM
  or in SM extensions through singlets,
  studied in Ref.~\cite{we}.
  %

\begin{widetext}

\begin{figure}[!]
\begin{center}
\includegraphics[width=1.0\textwidth]{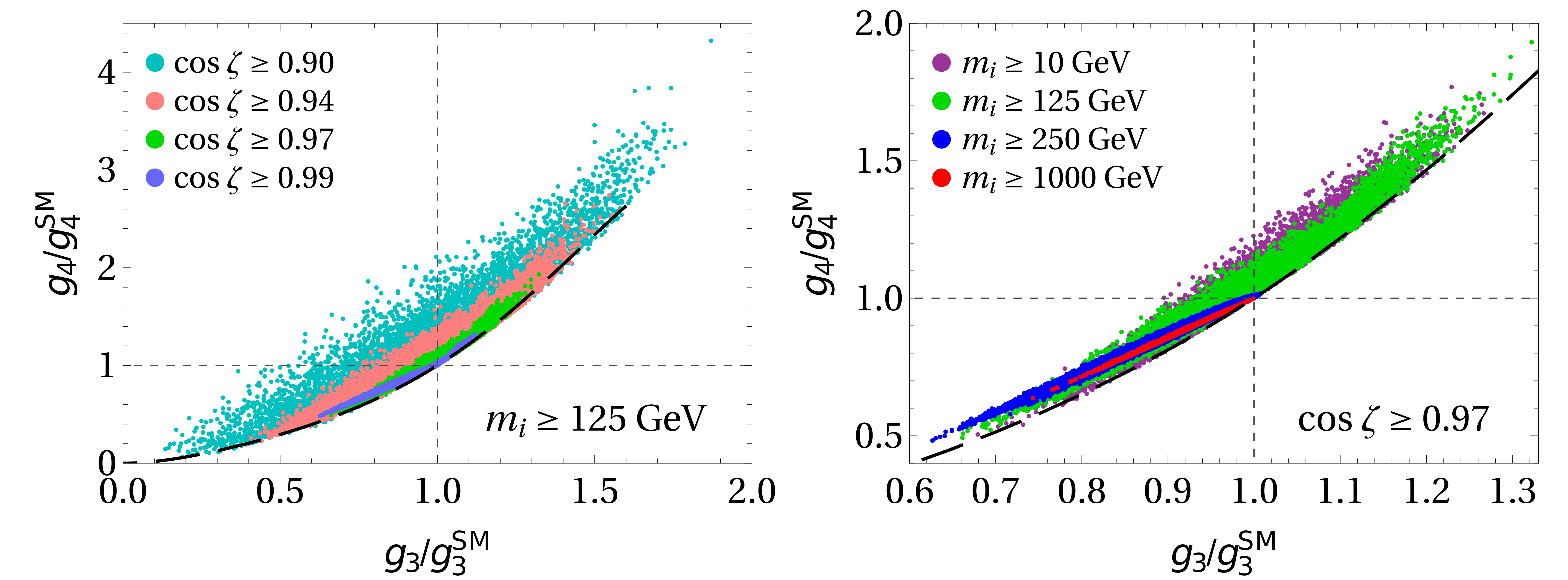}
\end{center}
\caption{Scatter plot of $g_4 \left/ g_4^\mathrm{SM} \right.$ \textit{versus}\/
  $g_3 \left/ g_3^\mathrm{SM} \right.$.
  The left panel compares the allowed regions
  for various values of $\cos{\zeta}$,
  when all the scalar masses are larger than 125\,GeV.
  The right panel compares the allowed regions
  for different lower bounds on the scalar masses.
  when $\cos{\zeta}$ is larger than 0.97.
  The straight dashed lines indicate the values of $g_3$ and $g_4$ in the SM.
  The long-dashed line is the same as the red dashed line
  in Fig.~\ref{fig:g34_models}.
  \label{fig:g34_CP-cons}}
\end{figure}

\end{widetext}
\item Scalar masses have a strong impact on $g_3$ and $g_4$.
  While for masses up to 125~GeV the ranges of the Higgs couplings
  remain unchanged,
  for larger masses $g_3 / g_3^\mathrm{SM}$ and $g_4 / g_4^\mathrm{SM}$
  must be below one.
\end{itemize}

The various constraints applied in this work
restrict deviations in $g_3$ and $g_4$ more strongly than
in other authors' works.
The analysis in Ref.~\cite{Durieux:2022hbu},
using an effective field theory approach,
predicts that deviations in $g_3$ could exceed 200\%.
In Ref.~\cite{kannikerecent} the bounds on $g_3$
are relaxed up to 35\% compared to in our study. 

Last but not least,
it is worth reminding the reader that $g_3$ suffers,
even in the SM,
very strong loop-level corrections,
as discussed in Ref.~\cite{Bahl:2022jnx}.
Therefore,
the tree-level analyses,
wherein fermions and their Yukawa couplings to the scalars are absent,
should be taken just as a first step
in the full study of the model with a scalar $SU(2)$ quadruplet.

\begin{acknowledgments}
D.J.\ thanks Jayita Lahiri and Thomas Biek\"otter
for communications on several aspects related to {\tt EVADE};
L.L.\ thanks Jo\~ao Viana for instruction on the relative merits
of {\tt EVADE} and {\tt BSMPT}.
The work of D.J.\ has been carried out
in the framework of the agreement of Vilnius University
with the Lithuanian Research Council No.~VS-13.
The work of L.L.\ has been supported by the Portuguese
Foundation for Science and Technology through projects UIDB/00777/2020,
UIDP/00777/2020,
CERN/FIS-PAR/0002/2021,
and CERN/FIS-PAR/0019/2021.
\end{acknowledgments}

\begin{appendix}

\setcounter{equation}{0}
\renewcommand{\theequation}{A\arabic{equation}}

\section{Derivation of the BFB conditions}
\label{BFBapp}

Since $\lambda_6$,
$\lambda_7$,
and $\lambda_8$ are assumed to be real,
one may write,
instead of Eq.~\eqref{V4},
\ba
\label{V4'}
V_4 &=& \frac{\lambda_1}{2}\, F_1^2
+ \frac{\lambda_2}{2}\, F_2^2
+ \lambda_3 F_1 F_2
+ \lambda_4 F_4
+ \lambda_5 F_5
\no &&
+ \sum_{p=6}^8
\lambda_p\, \mathrm{Re}\, \mathcal{F}_p.
\ea
Taking into account Eqs.~\eqref{tri} and~\eqref{prods},
one may define three vectors in $\mathbbm{R}^6$:
%
%
%
\bs
\label{vecs}
\begin{align}
\vec v_1 \equiv& \left( \begin{array}{c}
  \mathrm{Re} \left( a^2 \right) \\
  \mathrm{Im} \left( a^2 \right) \\
  \sqrt{2}\ \mathrm{Re} \left( a b \right) \\
  \sqrt{2}\ \mathrm{Im} \left( a b \right) \\
  \mathrm{Re} \left( b^2 \right) \\
  \mathrm{Im} \left( b^2 \right)
\end{array} \right),
\end{align}
\begin{align}
\vec v_2 \equiv& \left( \begin{array}{c}
  \mathrm{Re} \left( \sqrt{3}\, b c - a d \right) \\
  \mathrm{Im} \left( \sqrt{3}\, b c - a d \right) \\
  \sqrt{2}\ \mathrm{Re} \left( b d - a e \right) \\
  \sqrt{2}\ \mathrm{Im} \left( b d - a e \right) \\
  \mathrm{Re} \left( b e - \sqrt{3}\, a f \right) \\
  \mathrm{Im} \left( b e - \sqrt{3}\, a f \right)
\end{array} \right),
\end{align}
\begin{align}
\vec v_3 \equiv& \left( \begin{array}{c}
  \mathrm{Re} \left( d^2 - \sqrt{3}\, c e \right) \\
  \mathrm{Im} \left( d^2 - \sqrt{3}\, c e \right) \\
  \mathrm{Re} \left( d e - 3 c f \right) \left/ \sqrt{2} \right. \\
  \mathrm{Im} \left( d e - 3 c f \right)  \left/ \sqrt{2} \right. \\
  \mathrm{Re} \left( e^2 - \sqrt{3}\, d f \right) \\
  \mathrm{Im} \left( e^2 - \sqrt{3}\, d f \right)
\end{array} \right).
\end{align}
\es
Then,
Eq.~\eqref{V4'} becomes
\ba
\label{V4''}
V_4 &=&
\frac{\lambda_1}{2}\, F_1^2
+ \frac{\lambda_2}{2}\, F_2^2
+ \lambda_3 F_1 F_2
+ \lambda_4 F_4
+ \lambda_5 F_5
\no &&
+ \lambda_6\, \vec v_1 \cdot \vec v_2
+ \lambda_7\, \vec v_2 \cdot \vec v_3
+ \lambda_8\, \vec v_1 \cdot \vec v_3.
\ea
The squared magnitudes of the three vectors are
\bs
\label{mods}
\ba
\left| \left| \vec v_1 \right| \right|^2 &=& F_1^2,
\\
\left| \left| \vec v_2 \right| \right|^2 &=& \frac{3 F_1 F_2 - F_4}{2},
\\
\left| \left| \vec v_3 \right| \right|^2 &=& \frac{F_5}{2},
\ea
\es
respectively.

Boundedness from below means that $V_4$ must never be negative.
Therefore,
in order to derive the BFB conditions it is necessary to minimize $V_4$.

\subsection{Minimization of $V_4$ relative to the angles} \label{mini}

It is clear from Eqs.~\eqref{V4''} and~\eqref{mods} that $V_4$
depends only on $F_1$,
$F_2$,
$F_4$,
$F_5$,
and
\begin{itemize}
\item $\alpha$,
  \textit{viz.}\ the angle between $\vec v_1$ and $\vec v_2$;
\item $\beta$,
  \textit{viz.}\ the angle between $\vec v_1$ and $\vec v_3$;
\item $\gamma$,
  \textit{viz.}\ the angle between $\vec v_2$ and $\vec v_3$.
\end{itemize}
We firstly minimize $V_4$ relative to these three angles.
Geometric intuition tells us
(and we have confirmed it through explicit numerical examples)
that the minimum of $V_4$ is attained when the three vectors $\vec v_1$,
$\vec v_2$,
and $\vec v_3$ are co-planar in six-space,
\textit{i.e.}\ when they all lie on the same plane.
When that happens,
$\gamma$ is equal to $\alpha - \beta\ \mbox{mod}\ \pi$.

It is clear from Eq.~\eqref{V4''} that,
by changing the sign of any of the three vectors,
one may change the signs of two out of the three coefficients $\lambda_6$,
$\lambda_7$,
and $\lambda_8$,
but the product $\lambda_6 \lambda_7 \lambda_8$ remains invariant;
thus,
defining
\ba
\label{stu}
s & \equiv & \left| \lambda_6 \right|
\times \left| \left| \vec v_1 \right| \right|
\times\left| \left| \vec v_2 \right| \right|,
\no 
t & \equiv & \left| \lambda_8 \right|
\times \left| \left| \vec v_1 \right| \right|
\times\left| \left| \vec v_3 \right| \right|,
\no 
u & \equiv & \left| \lambda_7 \right|
\times \left| \left| \vec v_2 \right| \right|
\times\left| \left| \vec v_3 \right| \right|,
\ea
(note that $s$,
$t$,
and $u$ are \emph{by definition} positive)
there are two possibilities:
\begin{itemize}
\item either $\lambda_6 \lambda_7 \lambda_8$ is positive,
  and then one may re-define $\alpha$ and $\beta$ so that
  \bs
  \label{V4'''}
  \begin{align}
  V_4 =&
  \,\frac{\lambda_1}{2}\, F_1^2
  + \frac{\lambda_2}{2}\, F_2^2
  + \lambda_3 F_1 F_2 + \lambda_4 F_4 + \lambda_5 F_5
  \\  &
  + s \cos{\alpha} + t \cos{\beta} + u \cos{\left( \alpha - \beta \right)}
  \\ \equiv& \,V_4^{(1)};
  \end{align}
  \es
\item or $\lambda_6 \lambda_7 \lambda_8$ is negative,
  and then one may re-define $\alpha$ and $\beta$ so that
  \bs
  \label{V4''''}
  \begin{align}
  V_4 =&
  \,\frac{\lambda_1}{2}\, F_1^2
  + \frac{\lambda_2}{2}\, F_2^2
  + \lambda_3 F_1 F_2 + \lambda_4 F_4 + \lambda_5 F_5
  \\ & 
  - s \cos{\alpha} - t \cos{\beta} - u \cos{\left( \alpha - \beta \right)}
  \\ \equiv&
  \,V_4^{(2)}.
  \end{align}
  \es
\end{itemize}

The minimization of $V_4^{(1)}$ relative to $\alpha$ and $\beta$ proceeds
according to a line explained in Refs.~\cite{branco,book}.
One firstly defines
\be
\label{lambda0}
\lambda \equiv
- s^4 t^4 - s^4 u^4 - t^4 u^4
+ 2 s^2 t^2 u^2 \left( s^2 + t^2 + u^2 \right),
\ee
and
\be
\label{lambdah}
\hat{\lambda} \equiv
s \cos{\alpha} + t \cos{\beta} + u \cos{\left( \alpha - \beta \right)}.
\ee
There are then two possibilities:
\begin{itemize}
\item either $\lambda < 0$,
  and then the extrema of $V_4^{(1)}$
  occur when both $\alpha$ and $\beta$ are either $0$ or $\pi$,
  with
  \be
  \label{vcpfpf}
   \hat{\lambda} =
  \mbox{either}\ s - t - u\ \mbox{or}\ t - u - s\ \mbox{or}\ u - s - t;
  \ee
\item or $\lambda > 0$,
  and then there is an extremum of $V_4^{(1)}$ where
  \be
  \label{nhg}
  \hat{\lambda}
  = \frac{- s^2 t^2 - s^2 u^2 - t^2 u^2}{2 s t u},
  \ee
  which is \emph{smaller} than $s - t - u$,
  $t - u - s$,
  and $u - s - t$.
\end{itemize}

Thus,
there are in total three possibilities:
\begin{enumerate}
\item If $\lambda_6 \lambda_7 \lambda_8 > 0$ and $\lambda \le 0$,
  then the minimum of $V_4^{(1)}$ occurs when two
  out of the three angles $\alpha$,
  $\beta$,
  and $\alpha - \beta$ are equal to $\pi$
  and the third angle is equal to $0$,
  \textit{viz.}\ when
  \be
  \label{min1}
  \hat{\lambda}
  = \left\{ \begin{array}{lcl}
    s - t - u &\Leftarrow& \mathrm{min} \left( s,\, t,\, u \right) = s, \\
    t - s - u &\Leftarrow& \mathrm{min} \left( s,\, t,\, u \right) = t, \\
    u - s - t &\Leftarrow& \mathrm{min} \left( s,\, t,\, u \right) = u.
  \end{array} \right.
  \ee
  It is important to note that the three values
  of $s \cos{\alpha} + t \cos{\beta} + u \cos{\left( \alpha - \beta \right)}$
  in Eq.~\eqref{min1} \emph{can always occur},
  irrespective of the sign of $\lambda$.
  Therefore,
  \bs
  \label{oiu}
  \begin{align}
  &\frac{\lambda_1}{2}\, F_1^2 + \frac{\lambda_2}{2}\, F_2^2
  + \lambda_3 F_1 F_2 + \lambda_4 F_4 + \lambda_5 F_5 \no 
  & + u - s - t \ge 0,
  \\
  &\frac{\lambda_1}{2}\, F_1^2 + \frac{\lambda_2}{2}\, F_2^2
  + \lambda_3 F_1 F_2 + \lambda_4 F_4 + \lambda_5 F_5 \no 
  & + t - s - u \ge 0,
  \\
  &\frac{\lambda_1}{2}\, F_1^2 + \frac{\lambda_2}{2}\, F_2^2
  + \lambda_3 F_1 F_2 + \lambda_4 F_4 + \lambda_5 F_5 \no 
  &+ s - t - u \ge 0
  \end{align}
  \es
  are necessary BFB conditions whenever $\lambda_6 \lambda_7 \lambda_8 > 0$;
  if moreover $\lambda \le 0$,
  then they are \emph{necessary and sufficent} BFB conditions.
\item if $\lambda_6 \lambda_7 \lambda_8 > 0$ and $\lambda > 0$,
  then Eq.~\eqref{nhg} gives the minimum of $V_4^{(1)}$,
  hence the necessary and sufficient BFB condition is
  \begin{align}
  \label{mfrjjif}
  &\frac{\lambda_1}{2}\, F_1^2 + \frac{\lambda_2}{2}\, F_2^2
  + \lambda_3 F_1 F_2 + \lambda_4 F_4 + \lambda_5 F_5 \no
  & - \frac{s t}{2 u} - \frac{s u}{2 t} - \frac{t u}{2 s} \ge 0.
  \end{align}
\item If $\lambda_6 \lambda_7 \lambda_8 < 0$,
  then the minimum of $V_4^{(2)}$ occurs when $\alpha = \beta = 0$.
  Thus,
  in that case
  \begin{align}
  \label{oiu2}
  &\frac{\lambda_1}{2}\, F_1^2 + \frac{\lambda_2}{2}\, F_2^2
  + \lambda_3 F_1 F_2 + \lambda_4 F_4 + \lambda_5 F_5 \no
  &- s - t - u \ge 0
  \end{align}
  is the necessary and sufficient BFB condition.
\end{enumerate}

\subsection{Minimization of $V_4$ relative to $F_1 \left/ F_2 \right.$}

Now that we have achieved the minimization of $V_4$ relative to the angles
among the six-vectors $\vec v_{1,2,3}$,
we must perform the minimization relative to $F_1$ and $F_2$,
that have mass-squared dimension,
and $F_4$ and $F_5$,
that have dimension mass to the fourth power.
We follow our recent paper~\cite{ourrecent}.
We use $F_2$ to normalize all the four quantities,
by defining the three dimensionless parameters\footnote{We
have made the interchange $x \leftrightarrow 1 - x$
between the definition of $x$ in Eq.~\eqref{x}
and the corresponding definition in Ref.~\cite{ourrecent}.
Moreover,
in Ref.~\cite{ourrecent} we have called $\gamma_5$
to the quantity that we now call $y$.}
\be
r \equiv \frac{F_1}{F_2}, \quad
x \equiv \frac{1}{2} - \frac{F_4}{6 F_1 F_2} \label{x}, \quad
y \equiv \frac{F_5}{5 F_2^2}.
\ee
The parameter $r$ has semi-infinite range $\left[ 0,\, + \infty \right[$.
The parameters $x$ and $y$ are bounded by $0 \le x \le 1$,
$0 \le y \le 9/20$,
and~\cite{ourrecent}
\be
5 y - 9 x \left( 1 - x \right) \le 0.
\ee
We perform analytically the minimization relative to $r$;
the minimization relative to $x$ and $y$ must be left to numerics.
We write
\be
\frac{s}{F_2^2} = s_0 r \sqrt{r}, \quad
\frac{t}{F_2^2} = t_0 r, \quad
\frac{u}{F_2^2} = u_0 \sqrt{r},
\ee
where---remember Eqs.~\eqref{mods} and~\eqref{stu}---
\be
s_0 = \sqrt{3 x} \left| \lambda_6 \right|, \quad
t_0 = \sqrt{\frac{5 y}{2}} \left| \lambda_8 \right|, \quad
u_0 = \sqrt{\frac{15 x y}{2}} \left| \lambda_7 \right|
\ee
do not depend on $r$.

Conditions~\eqref{oiu} may be written
\bs
\label{uvyf}
\begin{align}
&\frac{\lambda_1}{2}\, r^2
- s_0 r \sqrt{r}
+ \left[ \lambda_3 + 3 \lambda_4 \left( 1 - 2 x \right)
  - t_0 \right] r
\no &
+ u_0 \sqrt{r}
+ \frac{\lambda_2}{2} + 5 \lambda_5 y \ge 0,
\\
&\frac{\lambda_1}{2}\, r^2
- s_0 r \sqrt{r}
+ \left[ \lambda_3 + 3 \lambda_4 \left( 1 - 2 x \right)
  + t_0 \right] r
\no &
- u_0 \sqrt{r}
+ \frac{\lambda_2}{2} + 5 \lambda_5 y \ge 0,
\\
&\frac{\lambda_1}{2}\, r^2
+ s_0 r \sqrt{r}
+ \left[ \lambda_3 + 3 \lambda_4 \left( 1 - 2 x \right)
  - t_0 \right] r
\no &
- u_0 \sqrt{r}
+ \frac{\lambda_2}{2} + 5 \lambda_5 y \ge 0.
\end{align}
\es
If $\lambda_6 \lambda_7 \lambda_8 > 0$ and $\lambda \le 0$,
then conditions~\eqref{uvyf} must hold for all non-negative $\sqrt{r}$.
We employ theorem~2 of Ref.~\cite{quartic}:
in order that
\be
\label{mvolococ}
c_4 \left( \sqrt{r} \right)^4 + c_3 \left( \sqrt{r} \right)^3
+ c_2 \left( \sqrt{r} \right)^2 + c_1 \sqrt{r} + c_0
\ee
be non-negative for all positive $\sqrt{r}$,
one must require that\footnote{Since $0 \le y \le 9/20$,
the conditions~\eqref{uuu} are equivalent to conditions~\eqref{iop}.}
\be
\label{uuu}
c_4 \equiv \frac{\lambda_1}{2} > 0, \quad
c_0 \equiv \frac{\lambda_2}{2} + 5 \lambda_5 y > 0;
\ee
moreover,
after making the definitions~\eqref{alfabetagama} and~\eqref{DeltaLambdas},
condition~\eqref{poss} must hold.
If $\lambda_6 \lambda_7 \lambda_8 > 0$,
one must apply this theorem
to each of the three conditions~\eqref{uvyf} in succession,
thereby getting rid of the parameter $r$.

We next consider the quantity $\lambda$ of Eq.~\eqref{lambda0}.
One easily finds that
\be
\frac{\lambda}{F_2^{16}} = \frac{225\, r^6 x^2 y^2}{4}
\left( l_4 r^4 + l_3 r^3 + l_2 r^2 + l_1 r + l_0 \right),
\ee
with $r$-independent quantities $l_4, \ldots, l_0$ given by Eqs.~\eqref{theLs}.
On the other hand,
\be
\frac{- s^2 t^2 - s^2 u^2 - t^2 u^2}{2 s t u F_2^2}
= - \frac{s_0 t_0}{2 u_0}\, r^2 - \frac{s_0 u_0}{2 t_0}\, r
- \frac{t_0 u_0}{2 s_0}.
\ee
Therefore,
the quantity in Eq.~\eqref{mfrjjif} is equal to
\be
F_2^2 \left( p_2 r^2 + p_1 r + p_0 \right), \label{27a}
\ee
with $p_2$,
$p_1$,
and $p_0$ given by Eqs.~\eqref{kfjodp1}.
If $\lambda_6 \lambda_7 \lambda_8 > 0$ and $\lambda > 0$,
\textit{i.e.}\ when $l_4 r^4 + l_3 r^3 + l_2 r^2 + l_1 r + l_0 > 0$,
the quantity~\eqref{27a} must be non-negative.
In order to see where this leads us,
one firstly has to solve Eq.~\eqref{lll} for each input $\lambda_6$,
$\lambda_7$,
$\lambda_8$,
$x$,
and $y$.
Note that both $l_4$ and $l_0$ are negative,
\textit{cf.}\ Eqs.~\eqref{L4} and~\eqref{L0}.
Equation~\eqref{lll} has either zero,
two,
or four real and positive roots.
\begin{itemize}
\item If Eq.~\eqref{lll} has zero real and positive roots,
  then $\lambda$ is negative for every positive $r$.
\item If Eq.~\eqref{lll} has two real and positive roots---call them
  $r_1$ and $r_2$,
  with $r_2 > r_1$---then $\lambda > 0$ for $r_1 < r < r_2$
  and $\lambda \le 0$ for $r \le r_1$ and for $r \ge r_2$.
  One makes the change of variables $r \to m$,
  where
  \be
  m = \frac{r_1 - r}{r - r_2}.
  \ee
  Then,
  \be
  \label{cm9g9f0}
  r_1 < r < r_2\ \Leftrightarrow\ 0 < m < + \infty.
  \ee
  Furthermore,
%
\be
 \begin{array}{c}
  p_2 r^2 + p_1 r + p_0 \ge 0 \\
  \Updownarrow \\
  R_2 m^2 + 2 R_3 m + R_1 \ge 0,
\end{array} 
 \label{jbigfooo}
\ee
  where $R_1$,
  $R_2$,
  and $R_3$ are given by Eqs.~\eqref{kfjodp2}.
  According to~\eqref{cm9g9f0},
  the condition~\eqref{jbigfooo} must hold for all non-negative $m$.
  This is equivalent to conditions~\eqref{kfjodp3}~\cite{kannike}.
\item If Eq.~\eqref{lll} has four real and positive roots
  $r_1 < r_2 < r_3 < r_4$,
  then $\lambda > 0$ for either
$r_1 < r <r_2$ or $r_3 < r < r_4$; therefore,  
  both the conditions~\eqref{kfjodp3} and also the same conditions
  with $r_1 \to r_3$ and $r_2 \to r_4$ must hold.
\end{itemize}

\subsection{Minimization of $V_4$ relative to $x$ and $y$}

We have now got rid of the dependence of the BFB conditions on $r$.
The two parameters $x$ and $y$ remain.
Their domain is finite~\cite{ourrecent}:
\bs
\label{domain}
\begin{align}
& 0 \le y \le \frac{9}{20}, \\
& \frac{1}{2} - \frac{1}{3}\,  \sqrt{\frac{9}{4} - 5 y} \le x \le
\frac{1}{2} + \frac{1}{3}\, \sqrt{\frac{9}{4} - 5 y}.
\end{align}
\es
One needs to perform a numerical scan over $y$ and $x$ in this domain,
requiring that the BFB conditions hold
\emph{for every point} $\left( y, x \right)$ in the domain.
Our numerical analysis has shown that
it is sufficient to scan just the boundaries of the domain~\eqref{domain},
given by the straight line $y = 0$
and by the parabola $y = \left( 9/5 \right) x \left( 1 - x \right)$
for $x \in \left[ 0, 1 \right]$.
We have performed the scan with a step $\delta x = 0.001$,
which we have numerically verified to be an almost perfect approximation
to the continuous lines.

\setcounter{equation}{0}
\renewcommand{\theequation}{B\arabic{equation}}

\section{Uniqueness of the vacuum}
\label{uniq}

In this appendix we compute the value of the scalar potential (SP)
at some alternative extremum points.
The requirement that those extrema have a higher value of $V$
than the preferred vacua
places constraints on the parameters of the SP.
Notice that those constraints apply on both the quadratic and quartic
parts of $V$---named $V_2$ and $V_4$,
respectively,
in Eq.~\eqref{potV}---contrary to the unitarity and BFB constraints,
that only restrict $V_4$.
In practice,
we have applied both the analytical constraints derived in this appendix
and also the numerical package {\tt EVADE} to check that
each of our preferred vacua is indeed is the global minimum
of the corresponding SP.

\subsection{Simple alternative vacua}

We consider two alternative vacua:
the one where only $c$ has a VEV
(which is equivalent to only $f$ having a VEV),
and the one where both $c$ and $f$ have VEVs.
In the first case we obtain
\be
\left\langle 0 \left| V \right| 0 \right\rangle
= \mu_2^2 C + \frac{\lambda_2}{2}\, C^2,
\label{uqe1}
\ee
and in the second case we obtain
\be
\left\langle 0 \left| V \right| 0 \right\rangle
= \mu_2^2 \left( C + F \right) + \frac{\lambda_2}{2} \left( C + F \right)^2
+ 9 \lambda_5 C F.
\label{uqe2}
\ee
Equation~\eqref{uqe1} leads to
\be
C = - \frac{\mu_2^2}{\lambda_2} \ \Rightarrow \
\left\langle 0 \left| V \right| 0 \right\rangle
= - \frac{\left( \mu_2^2 \right)^2}{2 \lambda_2}.
\ee
Equation~\eqref{uqe2} leads to
\be
C = F = - \frac{\mu_2^2}{2 \lambda_2 + 9 \lambda_5} \ \Rightarrow \
\left\langle 0 \left| V \right| 0 \right\rangle
= - \frac{\left( \mu_2^2 \right)^2}{2 \lambda_2 + 9 \lambda_5}.
\ee
Therefore,
in order that the true vacuum be the one where both $b$ and $e$ have VEVs
and no other field has VEV,
one must require that
\bs
\label{alternative}
\ba
- \frac{\mu_2^2}{\lambda_2} > 0 &\Rightarrow&
V_0 < - \frac{\left( \mu_2^2 \right)^2}{2 \lambda_2};
\\
- \frac{\mu_2^2}{2 \lambda_2 + 9 \lambda_5} > 0 &\Rightarrow&
V_0 < - \frac{\left( \mu_2^2 \right)^2}{2 \lambda_2 + 9 \lambda_5}.
\ea
\es
Notice that,
since we already know about the BFB conditions~\eqref{iop},
\be
- \frac{\mu_2^2}{\lambda_2} > 0 \ \Leftrightarrow \
- \frac{\mu_2^2}{2 \lambda_2 + 9 \lambda_5} > 0 \ \Leftrightarrow \
\mu_2^2 < 0.
\ee

\subsection{Alternative solutions of Eqs.~\eqref{stab0} and~\eqref{stabpi}}
\label{subsec:Alternative}

Both Eqs.~\eqref{os1} and~\eqref{os3} lead to
\be
\tilde a T^3 + \tilde b T^2 + \tilde c T + \tilde d = 0,
\label{ur1}
\ee
where
\bs
\ba
\tilde a &=& \lambda_7^2,
\\
\tilde b &=& \left[ 6 \lambda_6 \lambda_7
  - 4 \left( \lambda_3 + \lambda_4 + \lambda_8 \right)^2 \right] U,
\\
\tilde c &=& \left[ 9 \lambda_6^2 U - 8 \left( \mu_1^2 + \lambda_1 U \right)
  \left( \lambda_3 + \lambda_4 + \lambda_8 \right) \right] U, \hspace*{3mm}
\\
\tilde d &=& - 4 \left( \mu_1^2 + \lambda_1 U \right)^2 U.
\ea
\es
Both Eqs.~\eqref{os2} and~\eqref{os4} lead to
\be
\tilde a' T^3 + \tilde b' T^2 + \tilde c' T + \tilde d' = 0,
\label{ur2}
\ee
where
\bs
\ba
\tilde a' &=& 4 \left( \lambda_2 + 4 \lambda_5 \right)^2,
\\
\tilde b' &=& 8 \left( \lambda_2 + 4 \lambda_5 \right)
\left[ \mu_2^2 + \left( \lambda_3 + \lambda_4 + \lambda_8 \right) U \right]
- 9 \lambda_7^2 U, \hspace*{12mm}
\\
\tilde c' &=&
4 \left[ \mu_2^2 + \left( \lambda_3 + \lambda_4 + \lambda_8 \right) U \right]^2
- 6 \lambda_6 \lambda_7 U^2,
\\
\tilde d' &=& - \lambda_6^2 U^3.
\ea
\es
Equations~\eqref{ur1} and~\eqref{ur2} produce
\be
T = \frac{\left( \tilde d  \tilde b'- \tilde b \tilde d' \right) \bar{b} + \bar{c} \bar{d}}{\left( \tilde b \tilde c' - \tilde c  \tilde b'\right) \bar{b} + \bar{b} \bar{d} + \bar{c}},
\label{t2}
\ee
and moreover there is a constraint:
\label{cons}
\begin{align}
& \frac{1}{\tilde a^2} \left\{ 
\tilde d^2 \bar{b}^3 
+ \left[ \tilde c^2 \bar{d}
  - \tilde c \tilde d \bar{c}
  - 2 \tilde b \tilde d \bar{d} \right] \bar{b}^2 
  \right. \no &
\phantom{\frac{1}{\tilde a^2}\{ } + \left[
  3 \tilde a \tilde d \bar{c} \bar{d}
  - \tilde b \tilde c \bar{c} \bar{d}
  + \tilde b^2 \bar{d}^2
  - 2 \tilde a \tilde c
  \bar{d}^2
  + \tilde b \tilde d \bar{c}^2
  \right] \bar{b} %
  \no & \left. 
\phantom{W..} + \, \tilde a^2 \bar{d}^3
 - \tilde a \tilde b \bar{c}
\bar{d}^2
 + \tilde a \tilde c ^2
\bar{d}  
 - \tilde a \tilde d \bar{c}^3
\right\} = 0,
\hspace*{7mm}
\end{align}
where $\bar{b} = \left (\tilde a \tilde b' - \tilde b \tilde a' \right)$,
$\bar{c} = \left( \tilde a \tilde c' - \tilde c \tilde a' \right)$, and
$\bar{d} = \left( \tilde a \tilde d' - \tilde d \tilde a' \right)$.
The quantity inside curly brackets in Eq.~\eqref{cons}
can be divided by $\tilde a^2$.
Equation~\eqref{cons} constitutes an order-nine polynomial equation for $U$,
which determines $U$;
afterwards,
Eq.~\eqref{t2} determines $T$ from $U$.
Of course,
only solutions with both positive $U$ and positive $T$ are acceptable,
but there may be more than one of them,
and one must compute the value of $V$ at each of them
and make sure that it is larger than the value of $V$ at the chosen vacuum.

\setcounter{equation}{0}
\renewcommand{\theequation}{C\arabic{equation}}

\section{The case $Y = 3/2$}
\label{Y32}

Let us suppose that the quadruplet $\Psi$ in Eq.~\eqref{multiplets1}
has weak hypercharge $3/2$ instead of $1/2$.
We multiply the $SU(2)$ doublet
\be
\tilde \Phi = \left( \begin{array}{c} b^\ast \\ - a^\ast
\end{array} \right)
\ee
by the quadruplet $\Psi$ to obtain the triplet
\be
\left( \tilde \Phi \otimes \Psi \right)_\mathbf{3}
= \frac{1}{\sqrt{3}} \left( \begin{array}{c}
  \sqrt{3}\, a^* c + b^* d \\
  \sqrt{2} \left( a^* d + b^* e \right) \\
  a^* e + \sqrt{3}\, b^* f
\end{array} \right),
\label{mbvlcpd}
\ee
that has weak hypercharge $1$.
Then,
\ba
\mathcal{F}_9 &\equiv&
\left( \tilde \Phi \otimes \Psi \right)_\mathbf{3}^\dagger
\left( \Phi \times \Phi \right)_\mathbf{3}
= \no
&& a^3 c^\ast + b^3 f^\ast + \sqrt{3}\, a b \left( a d^\ast + b e^\ast \right)
\ea
is gauge-invariant.
The quartic part of the scalar potential therefore is
\ba
\label{V4dois}
V_4 &=& \frac{\lambda_1}{2}\, F_1^2 + \frac{\lambda_2}{2}\, F_2^2
+ \lambda_3 F_1 F_2 + \lambda_4 F_4 + \lambda_5 F_5 \no
&&+ \left( \xi_6 \mathcal{F}_9 + \mathrm{H.c.} \right).
\ea

Referring to the triplet~\eqref{mbvlcpd},
notice that
\begin{align}
& \left| \sqrt{3}\, a^* c + b^* d \right|^2
+ 2 \left| a^* d + b^* e \right|^2 
+ \left| a^* e + \sqrt{3}\, b^* f \right|^2
\no &
= \frac{F_4 + 3 F_1 F_2}{2}
\label{ifofoof}
\end{align}
is gauge-invariant.

The unitarity conditions for $V_4$ in Eq.~\eqref{V4dois} are
\bs
\ba
\left| \lambda_2 \right| &<& M,
\\
\left| \lambda_2 + 3 \lambda_5 \right| &<& M,
\\
\left| \lambda_2 + 9 \lambda_5 \right| &<& M,
\\
\left| \lambda_2 + 10 \lambda_5 \right| &<& M,
\\
\left| \lambda_3 \right| + 3 \left| \lambda_4 \right| &<& M,
\\
\left| \lambda_3 - 5 \lambda_4 \right| &<& M,
\\
\left| \lambda_1 + \tilde Q_1 \right|
+ \sqrt{\left( \lambda_1 - \tilde Q_1 \right)^2
  + 96 \left| \xi_6 \right|^2} &<& 2 M, \hspace*{3mm}
\label{bg}
\\
\left| \lambda_1 + \tilde Q_2 \right|
+ \sqrt{\left( \lambda_1 - \tilde Q_2 \right)^2
  + 160 \lambda_4^2} &<& 2 M,
\\
\left| 3 \lambda_1 + \tilde Q_3 \right|
+ \sqrt{\left( 3 \lambda_1 - \tilde Q_3 \right)^2
  + 32 \lambda_3^2} &<& 2 M,
\ea
\es
where $\tilde Q_1 = \lambda_3 + 5 \lambda_4$, 
$\tilde Q_2 = \lambda_2 - 11 \lambda_5$,
$\tilde Q_3 =  5 \lambda_2 + 15 \lambda_5$, 
and $M = 8 \pi$.

Because of Eq.~\eqref{ifofoof},
the lowest possible value of $V_4$ is
\begin{align}
\left( V_4 \right)_\mathrm{minimum}
= & \; \frac{\lambda_1}{2}\, F_1^2 + \frac{\lambda_2}{2}\, F_2^2
+ \lambda_3 F_1 F_2 + \lambda_4 F_4 + \lambda_5 F_5
\no &
- \left| \xi_6 \right| F_1 \sqrt{\frac{2 \left( F_4 + 3 F_1 F_2 \right)}{3}}.
\end{align}
Therefore,
in order to assure that $V_4$ is BFB one must,
for every $x$ and $y$ in the range~\eqref{domain},
employ Theorem~2 of Ref.~\cite{quartic} using
\bs
\ba
c_4 &=& \frac{\lambda_1}{2},
\\
c_3 &=& - 2 \left| \xi_6 \right| \sqrt{1 - x},
\\
c_2 &=& \lambda_3 + 3 \lambda_4 \left( 1 - 2 x \right),
\\
c_1 &=& 0,
\\
c_0 &=& \frac{\lambda_2}{2} + 5 \lambda_5 y.
\ea
\es
The scan over $x$ and $y$ must be performed numerically.

The neutral fields are $b$ and $f$.
The potential for them two is
\begin{align}
V_{b,f} = &\; \mu_1^2 B + \mu_2^2 F
+ \frac{\lambda_1}{2}\, B^2
+ \frac{\lambda_2}{2}\, F^2
+ \left( \lambda_3 + 3 \lambda_4 \right) B F
\no &
+\left(  \xi_6 b^3 f^* +  \xi_6^* {b^*}^3 f \right).
\end{align}
Since there is only one term in $V_{b,f}$ that
is sensitive to the relative phase of $b$ and $f$,
there can be no $CP$ violation,
\textit{i.e.}
\be
\left( \xi_6 b^3 f^* + \xi_6^* {b^*}^3 f \right)_\mathrm{minimum}
= - 2 \left| \xi_6 \right| \sqrt{B^3 F}.
\ee

\begin{widetext}

\begin{figure}[!]
\begin{center}
\includegraphics[width=1.0\textwidth]{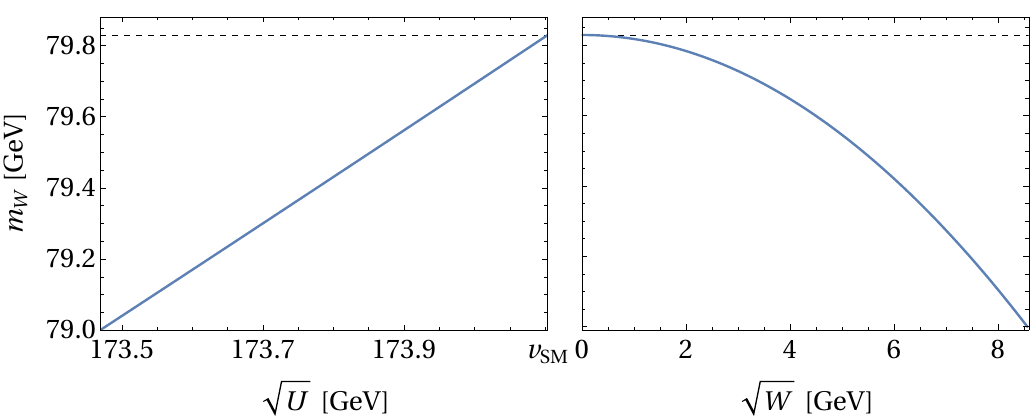}
\end{center}
\caption{The dependence of the VEVs $\sqrt{U}$ and $\sqrt{W}$
  on the mass $m_W$ according to Eq.~\eqref{eq:UW} in $Y = 3/2$ case.
  The black dashed line indicates the value of the mass $m_W^\mathrm{SM}$
  of the gauge bosons $W^\pm$ in the SM.
  \label{fig:VEVs_Y32}}
\end{figure}

\end{widetext}

Let the VEVs of $b$ and $f$ be $u$ and $w$,
respectively.
Let $U = \left| u \right|^2$ and $W = \left| w \right|^2$.
Then~\cite{albergaria},
\bs
\ba
m_W^2 &=& g^2\, \frac{U + 3 W}{2},
\\
m_Z^2 &=& \frac{g^2}{c_W^2}\, \frac{U + 9 W}{2}.
\ea
\es
The quantities $U$ and $W$ are determined from the input observables
through---remember Eq.~\eqref{vSM}---
\bs
\ba
U + 3 W &=& v_\mathrm{SM}^2,
\\
U + 9 W &=& \frac{m_Z^2 \left( \sqrt{2} G_F m_W^2
  - \pi \alpha_\mathrm{em} \right)}{\left( 2 G_F m_W^2 \right)^2}.
\ea
\label{eq:UW}
\es
The relation between the VEVs and $m_W$ is given in Figure~\ref{fig:VEVs_Y32}.

Notice that in this case one must choose the input observables $G_F$,
$\alpha_\mathrm{em}$,
$m_W^2$,
and $m_Z^2$ in such a way that
\be
\sqrt{2} G_F m_W^4 \le m_Z^2 \left( \sqrt{2} G_F m_W^2 - \pi \alpha_\mathrm{em}
\right).
\label{cond_Y32}
\ee
This is the opposite of condition~\eqref{condinput}
and \emph{contradicts} the recent measurement of $m_W$
by the CDF Collaboration~\cite{CDF}.

One has
\begin{align}
V_0 \equiv & \,\left\langle 0 \left| V_{b,f} \right| 0 \right\rangle
= \mu_1^2 U + \mu_2^2 W
+ \frac{\lambda_1}{2}\, U^2
+ \frac{\lambda_2}{2}\, W^2
\no &
+ \left( \lambda_3 + 3 \lambda_4 \right) U W
- 2 \left| \xi_6 \right| \sqrt{U^3 W}.
\end{align}
The extremum equations are
\bs
\label{29}
\ba
0 &=& \mu_1^2 + \lambda_1 U + \left( \lambda_3 + 3 \lambda_4 \right) W
- 3 \left| \xi_6 \right| \sqrt{U W}, \hspace*{7mm}
\label{29a}
\\
0 &=& \mu_2^2 + \lambda_2 W + \left( \lambda_3 + 3 \lambda_4 \right) U
- \left| \xi_6 \right| \sqrt{\frac{U^3}{W}}.
\label{29b}
\ea
\es
Therefore,
\be
V_0 = \frac{\mu_1^2 U + \mu_2^2 W}{2}.
\label{V0}
\ee
Conditions~\eqref{alternative} apply.

Let us define
\bs
\ba
x_1 &=& \lambda_1^2,
\\
x_2 &=& 2 \lambda_1 \left[ \mu_1^2 + \left( \lambda_3 + 3 \lambda_4 \right) W
  \right] - 9 \left| \xi_6 \right|^2 W, \hspace*{3mm}
\\
x_3 &=& \left[ \mu_1^2 + \left( \lambda_3 + 3 \lambda_4 \right) W \right]^2.
\ea
\es
Then,
Eq.~\eqref{29a} implies that
\be
x_1 U^2 + x_2 U + x_3 = 0.
\label{thex}
\ee
Similarly,
let us define
\bs
\ba
y_1 &=& - \left| \xi_6 \right|^2,
\\
y_2 &=& \left( \lambda_3 + 3 \lambda_4 \right)^2 W,
\\
y_3 &=& 2 \left( \lambda_3 + 3 \lambda_4 \right)
\left( \mu_2^2 + \lambda_2 W \right) W,
\\
y_4 &=& \left( \mu_2^2 + \lambda_2 W \right)^2 W.
\ea
\es
Then,
Eq.~\eqref{29b} implies that
\be
y_1 U^3 + y_2 U^2 + y_3 U + y_4 = 0.
\label{they}
\ee
Equations~\eqref{thex} and~\eqref{they} together mean that
\be
U = \frac{x_1 x_3 y_2 - x_2 x_3 y_1 - x_1^2 y_4}{x_2^2 y_1 + x_1^2 y_3
  - x_1 x_2 y_2 - x_1 x_3 y_1}.
\label{i49943}
\ee
Equation~\eqref{i49943} gives $U$ as a function of $W$;
while $W$ itself is given by
\ba
\label{fjbogo}
0 &=& x_1^3 y_4^2 + x_1^2 \left(
- x_2 y_3 y_4
+ x_3 y_3^2
- 2 x_3 y_2 y_4 \right)
\no & &
+ x_1 \left( x_2^2 y_2 y_4 + 3 x_2 x_3 y_1 y_4 - x_2 x_3 y_2 y_3
\right. \no && \left.
\phantom{WW} + x_3^2 y_2^2 - 2 x_3^2 y_1 y_3 \right)
\no & &
- x_2^3 y_1 y_4
+ x_2^2 x_3 y_1 y_3
- x_2 x_3^2 y_1 y_2
+ x_3^3 y_1^2, \hspace*{5mm}
\ea
which is a polynomial equation of order six for $W$.
In this way,
after one has obtained all the parameters of the potential,
one may check whether there are other solutions of Eqs.~\eqref{29}
for $U$ and $W$ and whether those other solutions yield,
through Eq.~\eqref{V0},
a lower $V_0$.

If one expands $b$ and $f$ as
\be
b = \sqrt{U} + \frac{R_b + i I_b}{\sqrt{2}}
\quad \mbox{and} \quad
f = \sqrt{W} + \frac{R_f + i I_f}{\sqrt{2}},
\ee
then there is a linear combination of $I_b$ and $I_f$
that is a pseudoscalar eigenstate of mass,
with squared mass
\be
m_P^2 = \left( U + 9 W \right) \sqrt{\frac{U}{W}} \left| \xi_6 \right|.
\label{eq:m_P}
\ee
On the other hand,
$R_b$ and $R_f$ have mass terms given by
\be
V = \cdots + \frac{1}{2}
\left( \begin{array}{cc} R_b, & R_f \end{array} \right)
\left( \begin{array}{cc} \hat M_{11} & \hat M_{12} \\
  \hat M_{12} & \hat M_{22} \end{array} \right)
\left( \begin{array}{c} R_b \\ R_f \end{array} \right),
\ee

where
\bs
\ba
\hat M_{11} &=& 2 \lambda_1 U - \frac{3 m_P^2 W}{U + 9 W},
\\
\hat M_{22} &=& 2 \lambda_2 W + \frac{m_P^2 U}{U + 9 W},
\\
\hat M_{12} &=& \left[
  2 \left( \lambda_3 + 3 \lambda_4 \right) - \frac{3 m_P^2}{U + 9 W} \right]
\sqrt{U W}.
\ea
\es
These matrix elements are related to the squared masses $M_h$ and $M_H$
of the physical scalars and to a mixing angle $\zeta$ through
\bs
\ba
\hat M_{11} &=& M_h \cos^2{\zeta} + M_H \sin^2{\zeta}, \\
\hat M_{22} &=& M_h \sin^2{\zeta} + M_H \cos^2{\zeta}, \\
\hat M_{12} &=& \left( M_h - M_H \right) \cos{\zeta}  \sin{\zeta}.
\ea
\es
In this extension of the SM there are,
besides,
one scalar with electric charge $3$ and squared mass
\be
m_{+++}^2 = 3 \left( - 2 \lambda_4 U + 3 \lambda_5 W \right)
+ \frac{m_P^2 U}{U + 9 W},
\label{eq:m+++}
\ee
one scalar with electric charge $2$ and squared mass
\be
m_{++}^2 = 2 \left( - 2 \lambda_4 U + 3 \lambda_5 W \right)
+ \frac{m_P^2 U}{U + 9 W},
\label{eq:m++}
\ee
and one scalar with electric charge $1$ and squared mass
\be
m_+^2 = \left( - 2 \lambda_4 + \frac{m_P^2}{U + 9 W} \right) v_\mathrm{SM}^2.
\label{eq:m+}
\ee
Notice that the model predicts
\be
3 m_{++}^2 - 2 m_{+++}^2 = \frac{m_P^2 U}{U + 9 W}.
\ee

The three- and four-Higgs couplings are given by
\bs
\ba
g_3 &=& \frac{1}{\sqrt{2}} \left\{
\left( \lambda_1 \sqrt{U} - \left| \xi_6 \right| \sqrt{W} \right) \cos^3{\zeta}
+ \lambda_2 \sqrt{W} \sin^3{\zeta} 
\right. \no & &
+ \left[ \left( \lambda_3 + 3 \lambda_4 \right) \sqrt{W}
  - 3 \left| \xi_6 \right| \sqrt{U} \right] \cos^2{\zeta} \sin{\zeta}
\no & & \left.
+ \left( \lambda_3 + 3 \lambda_4 \right) \sqrt{U} \cos{\zeta} \sin^2{\zeta}
\right\},
\\
g_4 &=& \frac{1}{8} \left[ \lambda_1 \cos^4{\zeta} + \lambda_2 \sin^4{\zeta}
  + 2 \left( \lambda_3 + 3 \lambda_4 \right) \cos^2{\zeta} \sin^2{\zeta}
\right. \no && \left.
  - 4 \left| \xi_6 \right| \cos^3{\zeta} \sin{\zeta} \right],
\ea
\label{eq:Y=32_g34}
\es
respectively.

\subsection{Procedure}
\label{sec:procedure2}

The procedure that we have followed
for the numerical computations was the following:
\begin{enumerate}
\item We have used the values of
  $m_Z$,
  $\alpha_\mathrm{em}$,
  and $G_F$
  in Eqs.~\eqref{theinputs},
  while we have let $m_W$ vary between
  its SM value $m_W^\mathrm{SM} = 79.829\,\mathrm{GeV}$
  and $79.6\,\mathrm{GeV}$.
  We have computed $U$,
  $W$,
  and $v_\mathrm{SM}$ by using Eqs.~\eqref{eq:UW} and~\eqref{vSM},
  respectively.
\item We have inputted $\lambda_2$,
  $\lambda_4$,
  $\lambda_5$,
  $\xi_6$,
  $M_h = \left( 125\,\mathrm{GeV} \right)^2$,
  and an angle $\zeta$ such that $\left| \cos{\zeta} \right| > 0.9$.
\item We determined $m_P^2$ through Eq.~\eqref{eq:m_P}.
\item We determined $m_{+++}^2$ through Eq.~\eqref{eq:m+++},
  $m_{++}^2$ through Eq.~\eqref{eq:m++},
  and $m_+^2$ through Eq.~\eqref{eq:m+}.
\item We determined $M_H$ through
  \be
  M_H = \frac{1}{\cos^2{\zeta}}
  \left( 2 \lambda_2 W + \left| \xi_6 \right| \sqrt{\frac{U^3}{W}}
  - M_h \sin^2{\zeta} \right).
  \ee
\item We determined $\lambda_1$ through
  \be
  \lambda_1 = \frac{1}{2 U} \left(M_h \cos^2{\zeta} + M_H \sin^2{\zeta}
  + 3 \left| \xi_6 \right| \sqrt{U W} \right).
  \ee
\item We determined $\lambda_3$ through
  \ba
  \lambda_3 &=& \frac{1}{2 \sqrt{U W}}
  \left[ \left( M_h - M_H \right) \cos{\zeta} \sin{\zeta}
    + 3 \left| \xi_6 \right| U \right]
  \no &&  
  - 3 \lambda_4  .
  \ea
\item We computed $\mu_1^2$ and $\mu_2^2$ through Eqs.~\eqref{29}.
\item We applied the conditions of Eq.~\eqref{alternative}.
\item We applied the procedure to test for vacuum stability
  described in Eqs.~\eqref{29} to~\eqref{fjbogo}.
\item We checked whether the UNI and BFB conditions on $\lambda_{1\mbox{--}5}$
  and $\xi_6$ hold.
\item We performed the vacuum-stability test with {\tt EVADE}
  and discarded the points which are not a global minimum of the SP.
\item We computed $g_3$ and $g_4$ through Eqs.~\eqref{eq:Y=32_g34}.
\end{enumerate}

\end{appendix}

\providecommand{\href}[2]{#2}\begingroup\raggedright\endgroup

 \end{document}